\documentclass{article}

\usepackage{PRIMEarxiv}

\usepackage[utf8]{inputenc} 
\usepackage[T1]{fontenc}    
\usepackage{hyperref}       
\usepackage{url}            
\usepackage{booktabs}       
\usepackage{amsfonts}       
\usepackage{nicefrac}       
\usepackage{microtype}      
\usepackage{lipsum}
\usepackage{fancyhdr}       
\usepackage{graphicx}       
\graphicspath{{media/}}     
\usepackage{longtable}
\usepackage{float}
\usepackage{threeparttable}
\usepackage{multirow}
\usepackage{hyperref}
\usepackage{amsmath, amssymb}
\usepackage{amsthm}
\usepackage[mathscr]{euscript}
\usepackage{algorithm, algorithmicx}
\usepackage[noend]{algpseudocode}
\newcommand{\Continue}{\State \textbf{continue}}
\newcommand{\Break}{\State \textbf{break}}
\usepackage[caption=false,font=scriptsize]{subfig}

\DeclareMathOperator*{\argmax}{argmax}
\DeclareMathOperator*{\argmin}{argmin}
\usepackage{alltt}  %

\makeatletter
\newcommand{\StatexIndent}[1][3]{%
  \setlength\@tempdima{\algorithmicindent}%
  \Statex\hskip\dimexpr#1\@tempdima\relax}
\makeatother

\usepackage{enumitem}

\algnewcommand\algorithmicforeach{\textbf{for each}}
\algdef{S}[FOR]{ForEach}[1]{\algorithmicforeach\ #1\ \algorithmicdo}
\title{Efficient Community Detection in Large-Scale Dynamic Networks Using Topological Data Analysis
}

\author{
  Wei Guo \\
  Department of Industrial \& Systems Engineering \\
  University of Washington\\
  Seattle, WA 98195\\
  \texttt{weig@uw.edu} \\
   \And
  Ruqian Chen \\
  Department of Mathematics \\
  University of Washington \\
  Seattle, WA 98105 \\
  \texttt{ruqian@uw.edu} \\
  \And
  Yen-Chi Chen \\
  Department of Statistics \\
  University of Washington \\
  Seattle, WA 98105 \\
  \texttt{yenchic@uw.edu} \\
  \And
  Ashis G. Banerjee \\
  Department of Industrial \& Systems Engineering \\
  Department of Mechanical Engineering \\
  University of Washington\\
  Seattle, WA 98195\\
  \texttt{ashisb@uw.edu} \\
}

\begin{document}
\maketitle

\begin{abstract}
In this paper, we propose a method that extends the persistence-based topological data analysis (TDA) that is typically used for characterizing shapes to general networks. We introduce the concept of the  community tree, a tree structure established based on clique communities from the clique percolation method, to summarize the topological structures in a network from a persistence perspective. Furthermore, we develop efficient algorithms to construct and update community trees by maintaining a series of clique graphs in the form of spanning forests, in which each spanning tree is built on an underlying Euler Tour tree. With the information revealed by community trees and the corresponding persistence diagrams, our proposed approach is able to detect clique communities and keep track of the major structural changes during their evolution given a stability threshold. The results demonstrate its effectiveness in extracting useful structural insights for time-varying social networks.
\end{abstract}


\section{Introduction}

The need to understand the dynamical and functional behavior of real-world networked systems has initiated extensive investigation of network structures over the past decade. Meanwhile, the topological analysis in the recent neuroscience studies has demonstrated that it is able to extract intrinsic information from neural networks that is practically impossible to extract using other less recent techniques of network theory~\cite{giusti2015clique,curto2017can}.

There is no universal definition of a community (a.k.a. cluster or cohesive subgroup) in network theory. A loose definition is that a community is a subgraph such that ``the number of internal edges is larger than the number of external edges"~\cite{fortunato2016community}. One pioneering work in community detection was proposed by Girvan and Newman in 2002~\cite{girvan2002community}. The authors developed an algorithm in which the communities were isolated by the successive removal of the identified inter-community edges. Since then, many new techniques, such as spin models, random walks, optimization, synchronization, as well as traditional clustering methods have been presented~\cite{reichardt2004detecting,zhou2003distance,newman2004fast,arenas2006synchronization,capocci2005detecting}.

Most of the aforementioned methods deliver standard partitions in which each vertex is assigned to a single community. However, vertices are often shared between communities in real-world networks. Therefore, detecting overlapping communities has received a lot of attention in the recent past. The first and the most popular algorithm is the clique percolation method (CPM) proposed by Palla et al.~\cite{palla2005uncovering}. It is based on the assumption that the internal edges of a community are likely to form cliques due to their high densities. Thus, a community is defined as a $k$-clique chain, i.e., a union of all the $k$-cliques that can be reached from each other through a series of adjacent $k$-cliques, where a $k$-clique refers to a maximal clique with $k$ vertices and two $k$-cliques are adjacent if they share $k-1$ vertices. In terms of implementation, the first step of this method is to find all the maximal cliques in the network. The second step is to generate the pairwise clique overlap matrix and extract the $k$-clique matrix from it. The extracted matrix represents the adjacency matrix of the cliques with size $k$. Overlapping communities are then detected by finding all the connected components in the adjacency matrix. 

In the CPM, $k$ is a predefined input parameter that may render information loss by leaving a considerable fraction of vertices and edges out of the communities. Moreover, without a prior structural information of the network, it is difficult for one to choose the value of $k$ to identify meaningful communities. To preserve the structural information as much as possible, we propose to convert a network to a topological representation in the form of a $k$-clique based community tree to summarize the community structure in the network at each order. The evolutionary analysis of the community structure will then be transformed to track the topological changes of the $k$-clique based community tree over time. To this end, as opposed to the CPM for static networks, we develop an incremental algorithm that allows us to update community trees with incremental changes in a network to keep track of evolving communities. 

Furthermore, we address the robustness of communities with respect to vertex and/or edge updates. In a fast-changing network, many of vertex and/or edge updates over a given time window may not lead to any significant change in the global community structure. Thus, it is unnecessary to incrementally update the evolving communities for each time window. For this reason, we adopt a metric defined from a previous study~\cite{chen2017note} to assess the topological change in community trees, and integrate it into the incremental algorithm to further improve the computational efficiency.

Following this framework, the rest of this paper is organized as follows. In Section~\ref{sec: stability_CT}, we first review the concept and stability of community trees. Section~\ref{sec: CG_and_ETT} outlines the implementation aspects for building and updating a community tree from a undirected, unweighted network. The overall algorithm, named as Dynamic CPM, is presented in Section~\ref{sec: dynamic_CPM}. Experimental results on a diverse collection of social network datasets are discussed in Section~\ref{sec: network_TDA_results} followed by a discussion of our findings in Section~\ref{sec: network_TDA_discussion}. 

\section{Community Tree}
\label{sec: stability_CT}

Establishing a filtration to track the evolution of topological features across the scales lies at the core of topological data analysis (TDA). Chen et al.~\cite{chen2017note} extends this idea to the topological characterization of structural features in the form of communities in a network. Specifically, a $k$-clique community or $k$-community for short, denoted by $\mathcal{C}_k$, is adopted from the definition of a community in the CPM, and $k$ is called as its order. Note that by this definition, any arbitrary network $G$ is a 1-community. Since a $k$-clique itself is a $(k-1)$-clique chain, a $k$-community is also a $(k-1)$-community. This property allows any $k$-community $\mathcal{C}_k$ to grow a nested sequence of communities across different orders such that $\mathcal{C}_k \subseteq \mathcal{C}_{k-1} \subseteq \cdots \subseteq \mathcal{C}_1=G$, which forms a filtration of $G$.

From a topological point of view, each nested sequence of communities records the evolution of the starting community as a connected component. Thus, the birth time of the connected component is defined as the order of the starting community. We say the starting community with a higher order is born earlier. When two connected components merge at a certain order, the merging order is then defined as the death time of the connected component with a later birth time. We set the death time of the last remaining connected component, i.e., the one with the earliest birth time, to be 1. The birth and death time of each connected component in a community tree is also encoded in a persistent diagram (PD). A connected component with a longer persistence implies that its starting community behaves more robust against the changes occurring in the network. 

\subsection{Stability of Community Tree}
\label{sec: stability_CT}

\begin{figure}
\centering
\includegraphics[width=0.8\columnwidth]{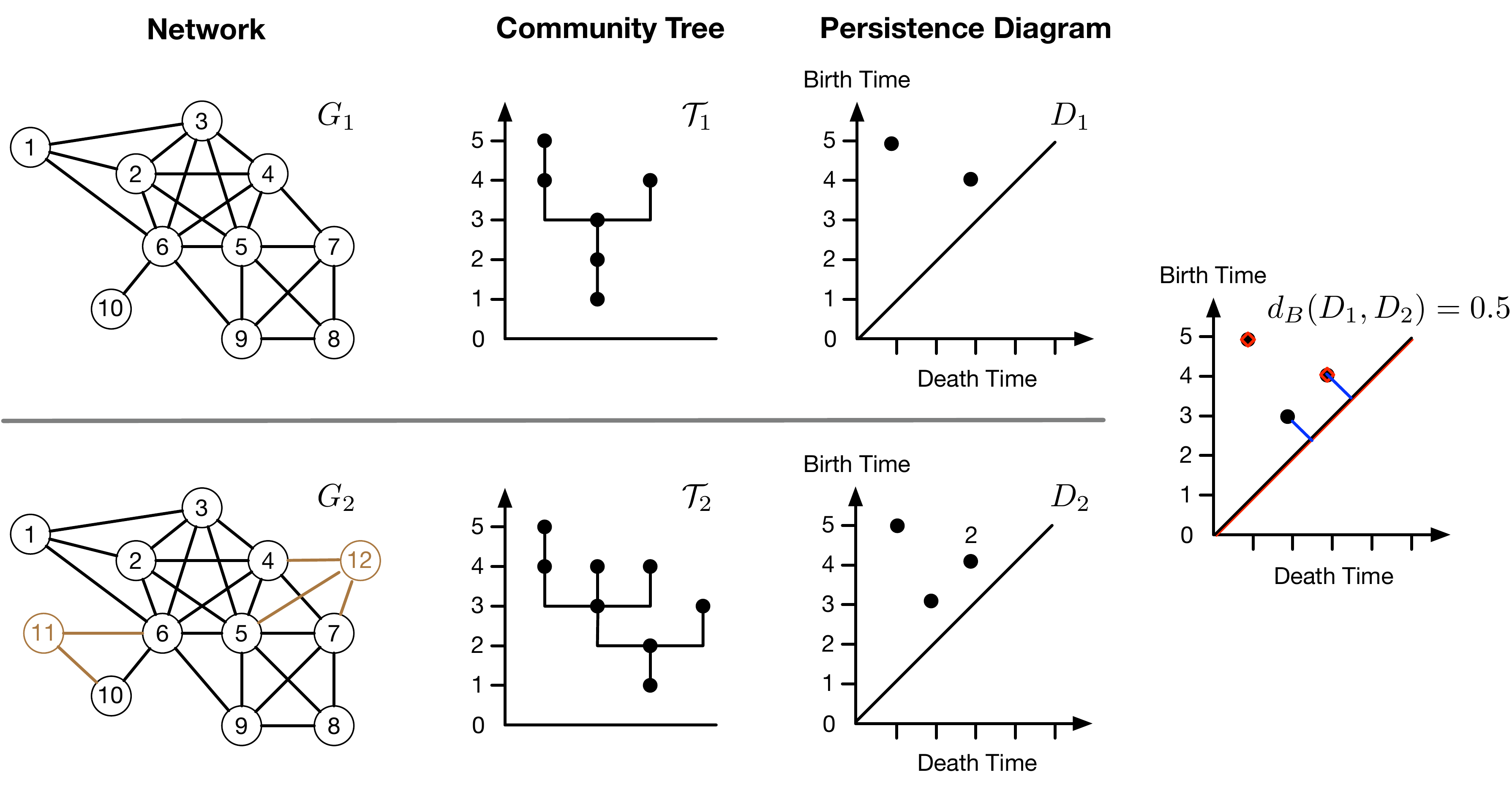}
\caption{The distance between two community trees. The community tree $\mathcal{T}_2$ and the corresponding PD $D_2$ are updated as $v_{11}$, $v_{12}$ and multiple edges are added (colored in brown) from $G_1$ to $G_2$. The PD in the right panel shows an optimal matching between the points in $D_1$ and $D_2$, which indicates $d_B(D_1, D_2) = 0.5$, i.e., $d_B(\mathcal{T}_1, \mathcal{T}_2) = 0.5$. On the other hand, $TSN = ASN =2$ since all the added edges are incident to $v_{11}$ and $v_{12}$.}
\label{fig_CT_stability}
\end{figure}

Chen et al.~\cite{chen2017note} also defines the bottleneck distance between two community trees as the bottleneck distance between their corresponding PDs, and proposes to use this metric to quantify how the two trees differ. To prove community trees are stable with respect to this metric, Chen et al. further introduces a quantity named {\it star number}. Mathematically, the {\it addition star number} ($ASN$) of $G_1$ and $G_2$ is given by 
$$ASN(G_2,G_1) = \min\{|V_0|: \nu(e)\cap V_0\neq\emptyset\,\,\, \forall e\in E(G_2)\backslash E(G_1)\},$$
where $V_0$ is a collection of vertices and $\nu(e)$ represents the two vertices of $e$. The {\it removal star number} ($RSN$) can be defined in a similar manner. The sum of $RSN$ and $ASN$ is the {\it total star number} ($TSN$). Figure~\ref{fig_CT_stability} provides an example of computing bottleneck distance of community trees and the $TSN$\footnote{The $x$ and $y$ axes are labeled as death time and birth time, respectively, in the PDs so that all the points corresponding connected components stay above the diagonal.}.

$TSN(G_2,G_1)$ attributes the change from $G_1$ to $G_2$ to a specified number of vertices. Moreover, it has been proved that the difference between two community trees is bounded above by
their $TSN$, i.e.,
$$
d_B(\mathcal{T}(G_1),\mathcal{T}(G_2)) \leq TSN(G_2,G_1). 
$$
This result guarantees that one can infer information about the changes to a community tree from the $TSN$. Thus, rather than directly update the community tree, we compute the $TSN$ first to estimate the topological change in community structures between the two networks.

\subsubsection{Algorithm for the upper bound of TSN}
\label{subsec: TSN-upper-bound}

Although computing the $TSN$ has been proved to be NP-complete in~\cite{chen2017note}, we can circumvent this issue by calculating an upper bound for the $TSN$. Given a temporal network for a duration of time, we usually take all the vertices and edges up to a particular time $t$ and create a graph $G_t$ for evolutionary analysis. Thus, the change from $G_{t-1}$ to $G_t$ is now an incremental case where vertices and edges are only added to $G_{t-1}$. Accordingly, computing the upper bound of $TSN$ is reduced to computing the upper bound of $ASN$.
\begin{algorithm}[htbp]
\caption{Compute an upper bound on the $TSN$ in the incremental case} 
\begin{algorithmic}[1]
\Procedure{TSN-upper-bound}{$G^\Delta_+$}
\State $\tau \gets |\Call{Find-VC}{G^\Delta_+}|$ \Comment{Use Bar-Yehuda and Even Algorithm to obtain $2$-OPT local ratio for minimum vertex cover.}
\State \Return $\tau$
\EndProcedure 
\end{algorithmic}
\label{alg::TSN}
\end{algorithm}

Here we employ an approximation algorithm to obtain the upper bound $\tau$ of $ASN$. As shown in Algorithm~\ref{alg::TSN}, we first compute an edge-difference graph $G^\Delta_+ = (V(G_t), E(G_t)\backslash E(G_{t-1}))$\footnote{$(V(G), E(G))$ and $(V(G_t), E(G_t))$ are henceforth denoted by $(V, E)$ and $(V_t, E_t)$, respectively, for simplicity.} as the input. Bar-Yehuda and Even's greedy algorithm~\cite{bar1981linear} is then used to find a vertex cover for $G^\Delta_+$. The solution given by this algorithm is guaranteed to be within 2 times the optimum solution. Hence, we use the size of this cover as an upper bound for $ASN$. The worst-case runtime for Bar-Yehuda and Even's algorithm is $O(\max{(|E_{t-1}|,|E_t|)})$. Therefore, Algorithm~\ref{alg::TSN} also has a worst case runtime of $O(\max{(|E_{t-1}|,|E_t|)})$.
\section{Clique Graph and Euler Tour Tree}
\label{sec: CG_and_ETT}

In this framework, we transform a network $G$ into a community tree through an auxiliary structure, termed as weighted {\it clique graph} (CG). In a weighted CG, each node represents an MC found in the network and the edge weight denotes the number of vertices shared by the two corresponding MCs. 
Since the presence of any single vertex in $G$ does not affect the resulting $\mathcal{T}(G)$, we ignore these MCs of size 1 when generating a weighted CG. Let $\mathcal{G}$ be a weighted CG that comprises all the MCs of size $s \geq 2$, and define $\mathcal{G}_i$ as the subgraph of $\mathcal{G}$ composed of CG nodes representing the MCs of size $s \geq i+1$ and edges with weights $w_{e_{\mathcal{G}}} \geq i$. Note that $\mathcal{G}_1 = \mathcal{G}$. 

\begin{figure}[!t]
\centering
\includegraphics[width=0.75\columnwidth]{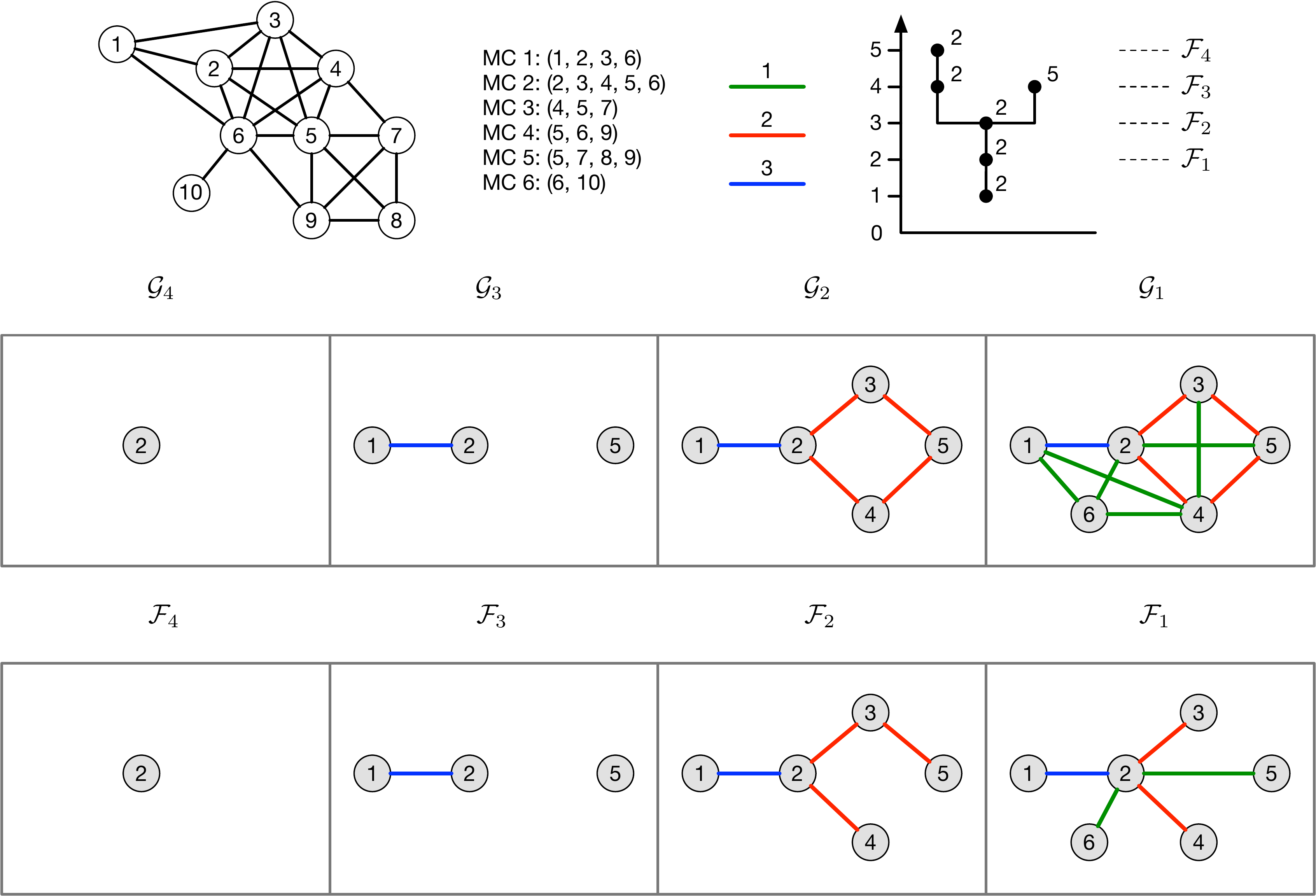}
\caption{An example of the sequence of weighted clique graphs $\mathcal{G}_i$ and the corresponding spanning forests $\mathcal{F}_i$, $i = \omega(G)-1, \dots, 1,$ for a given network $G$. The green, red and blue lines in $\mathcal{G}_i$ and $\mathcal{F}_i$ represent the edges with an edge weight of 1, 2 and 3, respectively. The connectivity information in $\mathcal{F}_i$ is summarized in the community tree $\mathcal{T}(G)$ at order $i+1$. For example, the representative CG nodes in $\mathcal{F}_3$ are 2 and 5, while the representative CG node in $\mathcal{F}_2$ is 2.}
\label{fig::CG}
\end{figure}

To perform the updates on the community tree efficiently as $G$ changes, we maintain a spanning forest (a spanning tree for each connected component (CC)) of $\mathcal{G}_i$, denoted by $\mathcal{F}_i$ for $i=1, \ldots, \omega(G)-1$. 
We refer to the edges in $\mathcal{F}_i$ as the {\it tree edges} and maintain an adjacency list of the tree edges for every node at each level $i$. Moreover, since the birth time of a CC in $\mathcal{F}_i$ equals the size of the largest MC included in the CC, we choose this MC as the {\it representative MC} for the CC and use it as the CC's ``label". If there are multiple MCs with the same maximum size, without loss of generality, the MC with the minimum ID is selected as the representative MC. Accordingly, we record the death time of the CC as the death time of the representative MC and use the ID of the representative MC as the CC's ID. We also name the CG node corresponding to the representative MC as {\it representative CG node}. See Figure~\ref{fig::CG} for an example.

\begin{figure}[!ht]
\centering
\includegraphics[width=0.75\columnwidth]{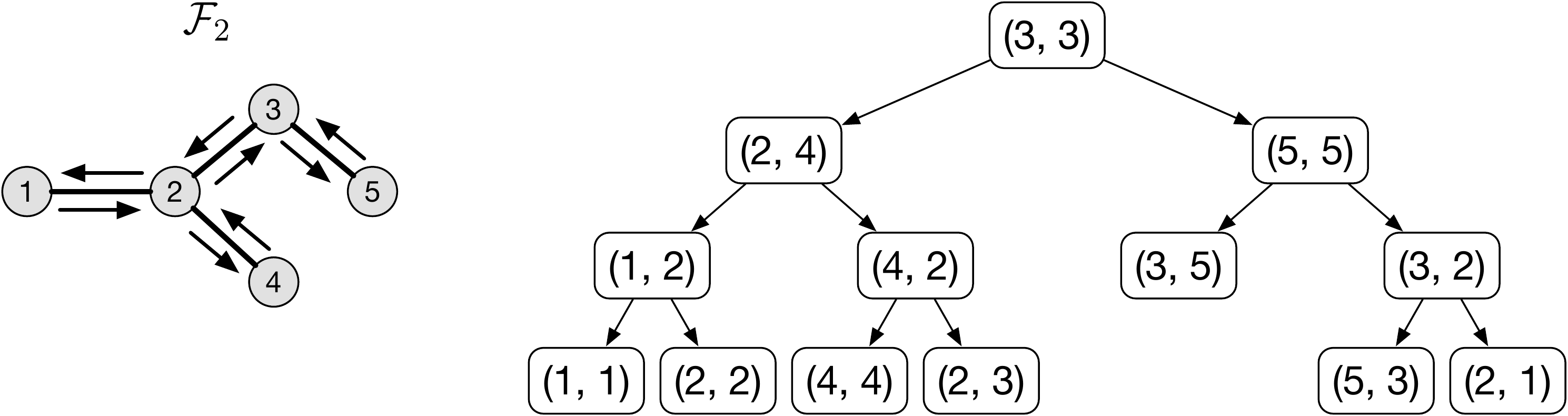}
\caption{An example of the Euler tour of a spanning tree in $\mathcal{F}_2$ (left) and one possible representation of this Euler tour as a balanced BST keyed by the index in the tour (right). In this example, the Euler tour is $(1,1)$-$(1,2)$-$(2,2)$-$(2,4)$-$(4,4)$-$(4,2)$-$(2,3)$-$(3,3)$-$(3,5)$-$(5,5)$-$(5,3)$-$(3,2)$-$(2,1)$. A spanning tree with $n$ nodes will be represented by a balanced BST with $2(n-1)+n$ nodes.}
\label{fig::ETT}
\end{figure}

Each spanning tree in the forest is built on an underlying {\it Euler Tour (ET) Tree} data structure, introduced by Henzinger and King \cite{henzinger1995randomized}. The ET tree is constructed based on the Euler tour of the spanning tree, which is essentially a depth-first traversal of the spanning tree and ends at the node at which it starts. Each ET tree is often stored as a balanced binary search tree (BST). The data structure was later modified by Tarjan \cite{tarjan1997dynamic} to better support operations on the tree nodes such as changing node values. Therefore, we adopt Tarjan's version of the data structure here. In this version, the Euler tour of a spanning tree is a sequence of arcs (directed edges) over the spanning tree with one ``loop" arc per node visited by the tour, i.e, each edge $(u,v)$ results in two arcs $(u,v)$ and $(v,u)$, while each node $v$ corresponds to a single loop arc $(v,v)$.  
See Figure~\ref{fig::ETT} for a simple illustration.

\begin{figure}[!ht]
\centering
\includegraphics[width=\columnwidth]{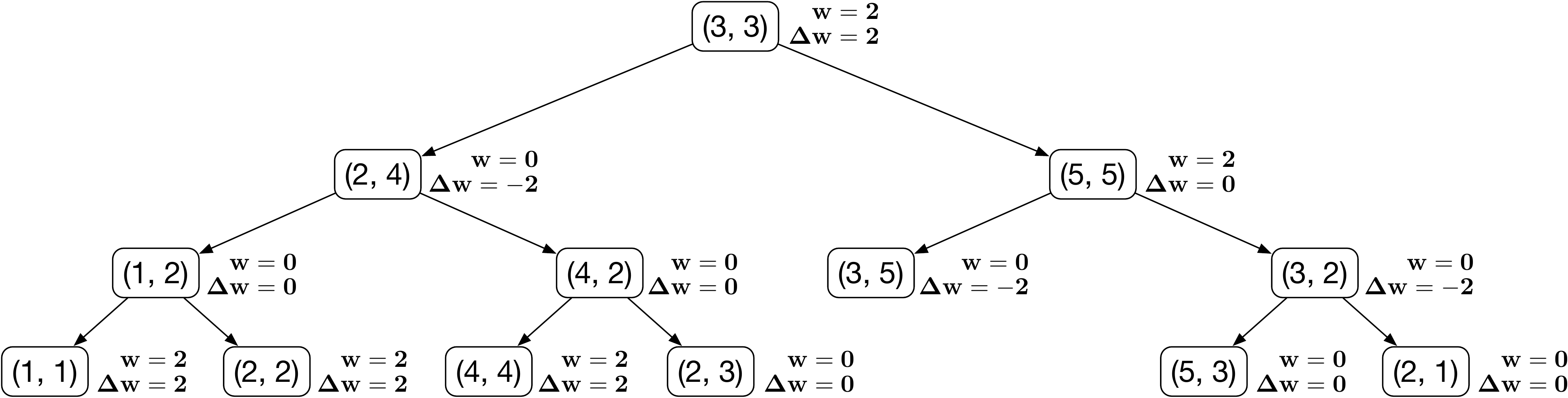}
\caption{An example of an ET tree for implementing the \textproc{Add-val} operation. It is sufficient to only maintain $\Delta w(x)$ for each node $x$ to find $w(x)$ in the ET tree.}
\label{fig::ETT_weight}
\end{figure}
We implement the Euler tour of a spanning tree with a splay tree \cite{sleator1985self}, which is a self-adjusting BST where a node is always splayed to the root through a series of rotations when accessed. 
Each node in the spanning tree holds a pointer to its corresponding node, referred to as {\it ET node}, in the splay tree. Particularly, the ET node corresponds to the representative CG node is referred to as {\it representative ET node}.
Meanwhile, each ET node in the splay tree also stores pointers to its parent, right and left child. With this representation, the following operations are supported in $O(\log n)$ amortized time using $O(n)$ space, where $n$ is the number of nodes in the spanning tree(s) involved in the operation \cite{tarjan1997dynamic}. 

\begin{itemize}[nosep]
\item \Call{Connected}{$u,v$}: Return if $u$ and $v$ are in the same spanning tree. 
\item \Call{Link}{$u, v$}: If $u$ and $v$ are in different spanning trees, insert an edge ($u, v$) connecting the trees together.
\item \Call{Cut}{$u, v$}: Delete the edge ($u, v$) from a spanning tree, splitting the tree into two trees.
\item \Call{Add-val}{$v, \alpha$}: Add $\alpha$ to the value of each node in a spanning tree containing node $v$.
\end{itemize}

\Call{Connected}{} is implemented by using the $\Call{Find-root}{}$ operation in the splay tree. Since linking two trees and cutting a tree each amounts to a fixed set of splitting and concatenation operations on Euler tours, \Call{Link}{} and \Call{Cut}{} are realized by a constant number of \Call{Split}{} and \Call{Join}{} operations of the splay tree \cite{tarjan1997dynamic}. 

To handle the \Call{Add-val}{} operation efficiently, we store the value for each node of the splay tree implicitly. Specifically, let $w(x)$ be the value associated with a splay tree node $x$. Rather than storing $w(x)$ at $x$, we store the difference $\Delta w(x)$ between $w(x)$ and the value of its parent, i.e.,
\begin{equation}
\Delta w(x) =
  \begin{cases}
    w(x)       & \quad \text{if } x \text{ is the root the splay tree}\\
    w(x)-w(p(x))  & \quad \text{if } x \text{ is a nonroot, where } p(x) \text{ is the parent of } x 
  \end{cases}
  \label{eq::delta-w}
\end{equation}
When $x$ is the root, then $w(x)$ itself is assigned to $\Delta w(x)$. In this manner, for each node $x$, $w(x)$ is computed by summing $\Delta w$ over all the ancestors of $x$. It also implies that adding $\alpha$ to $\Delta w(x)$ amounts to adding $\alpha$ to the values of all the descendants of $x$. Thus, \Call{Add-val}{$v, \alpha$} is realized by simply adding $\alpha$ to the node value $\Delta w$ of the root node of the splay tree that $v$'s corresponding ET node belongs to. $\Delta w(x)$ is updated in each rotation in $O(1)$ time during a splay step. One can check~\cite{sleator1985self} for details.

In our case, for a loop node, $w(x)$ is defined as the ID of the representative ET node of the ET tree that the loop node belongs to. This affiliation links each CG node in a spanning tree to the corresponding representative CG node, which facilitates the need to update the representative CG node in the new spanning tree whenever a tree edge is inserted between two CG nodes. For a nonloop node, $w(x)$ is assigned an arbitrary value of 0 since it can be simply ignored. An example is given in Figure~\ref{fig::ETT_weight} and more details can found in Algorithm~\ref{alg::update-rep}.
\section{Dynamic CPM}
\label{sec: dynamic_CPM}

We now present the overall method of our paper, which we call the Dynamic CPM. As laid out in Algorithm~\ref{alg::dynamicCPM}\footnote{The notations for Algorithm~\ref{alg::dynamicCPM} and the following algorithms are given in Table~\ref{table_symbols}.}, it essentially computes the sequence of community trees for a given sequence of complex networks that vary over time. We then directly obtain the network communities simply by scanning through each order of the computed trees one-by-one. 
\begin{algorithm}[!htbp]
\caption{Calculate sequence of community trees for dynamic networks that are represented as time-varying undirected, unweighted graphs $G_0, G_1, \ldots, G_T$}
\begin{algorithmic}[1]
\Procedure{Dynamic-CPM}{$G_0, G_1, \ldots, G_T, l$} 
\State $\mathscr{T}_0, \mathscr{M}_0, \mathcal{T}_0, m_0 \gets$ \Call{Build-CT}{$G_0$} \Comment{Algorithm~\ref{alg::buildCT}}
\State $Q \gets \emptyset$, $S\gets 0$, $t' \gets 0$
\For {$t = 1, \ldots, T$} 
\State $G^\Delta_+ \gets (V_t, E_t\backslash E_{t-1})$ 
\State $\tau_t \gets \Call{TSN-upper-bound}{G^\Delta_+}$ \Comment{Algorithm~\ref{alg::TSN}}
\If {$t \leq l$} 
\State \Call{Push-back}{$Q, \tau_t$}
\State $S \gets S+\tau_t$
\Else

\If {$\tau_t \geq S/l$} 
\State $V^\Delta_+ \gets V_t\backslash V_{t'}$, $E^\Delta_+ \gets E_t\backslash E_{t'}$
\State \scalebox{0.98}[1]{$\mathscr{T}_t, \mathscr{M}_t, \mathcal{T}_{t}, m_{t} \gets \Call{Update-CT}{G_{t'},V^\Delta_+, E^\Delta_+, \mathscr{T}_{t'}, \mathscr{M}_{t'}, \mathcal{T}_{t'}, m_{t'}}$} \Comment Algorithm~\ref{alg::update-CT}
\State $t' \gets t$, $\mathscr{T}_{t'} \gets \mathscr{T}_t$, $\mathscr{M}_{t'} \gets\mathscr{M}_t$, $m_{t'} \gets m_t$
\Else
\State $\mathcal{T}_t \gets \mathcal{T}_{t-1}$ \Comment{Retain previous community tree as the network has not changed substantially (from a topological perspective)}

\EndIf
\State $\tau_{t-l} \gets \Call{Pop-front}{Q}$
\State \Call{Push-back}{$Q, \tau_t$}
\State $S \gets S-\tau_{t-l}+\tau_t$
\EndIf
\EndFor 
\State \Return $\mathcal{T}_0, \mathcal{T}_1, \ldots, \mathcal{T}_T$ 
\EndProcedure
\end{algorithmic}
\label{alg::dynamicCPM}
\end{algorithm}

There are two distinct phases in the Dynamic CPM: \textit{initialization} and \textit{update if necessary}. When a graph, corresponding to a previously non-analyzed network, is fed to the method for the first time, Algorithm~\ref{alg::buildCT} is used to yield the initial community tree. Subsequently, as new graphs, which are  modified forms of the initial graph with vertex and edge insertions, are provided as inputs, we first use our TSN bound result on the bottleneck distance between any two successive graphs to decide whether new community trees should be computed. For efficiency, we can use  Algorithm~\ref{alg::TSN} to approximate the TSN bound. If we decide to compute new community trees, then Algorithm~\ref{alg::update-CT} is used to update the trees without recomputing them from scratch. We present the initial community tree construction algorithm and the update algorithm (along with all their sub-routines) in the Appendix. 

It is worth noting the importance of the update condition (line 11) in Algorithm~\ref{alg::dynamicCPM}. We are typically interested only in communities with strong (topological) persistence rather than communities that are short-lived and change even for minor modifications to the networks. In fact, we do not want to update the detected communities in such scenarios. Therefore, the choice of update condition provides a measure of control over the persistence of the communities in addition to rendering our community tracking method more efficient. Here, we update $\mathcal{T}_t$ if $\tau_t$ is no less than a moving average of length $l$. In the following experiments, the length $l$ is selected to be 3.

Figure~\ref{fig_sms_evolution} provides a realistic scenario where the Dynamic CPM works well. In this example,
we recompute the community trees at $t_4$ and $t_7$ because their respective TSN bounds are greater than the moving averages from previous windows. Indeed, the change from $G_0$ to $G_4$ is non-trivial as a new 5-community is born and a large number of lower order communities emerge. From $G_4$ to $G_7$, three more 5-communities appear along with more lower order communities being created. Thus, it makes sense to recompute the community tree. On the other hand, taking the set of 3-communities as an example, the unchanged community trees at order 3 well approximate the real 3-communities found by the CPM as is shown in Figure~\ref{fig_network_acc}.

\section{Results}
\label{sec: network_TDA_results}
We first briefly describe the network datasets that are used to evaluate our algorithm below. Although all the datasets are originally collected in the format of (sender, receiver, timestamp) tuples as directed networks, directionality is ignored in our analysis.
\begin{itemize}[nosep]
    \item Email networks [email-Enron~\cite{klimt2004enron}; email-Eu-core~\cite{paranjape2017motifs}]: Each vertex is an email address and an edge $(v_i, v_j)$ means that person $v_i$ sent or received an e-mail to or from person $v_j$. email-Enron records the communication among Enron employees mostly from November 1998 to June 2002\footnote{This dataset is downloaded from \url{http://www.cis.jhu.edu/~parky/Enron/}. The other datasets can be found from the links in \url{http://snap.stanford.edu/temporal-motifs/data.html}}, while email-Eu-core contains email data over 1 year from members of four different departments at a large European research institution.  
    \item Bitcoin network [bitcoin-au~\cite{kondor2014rich}]: Each vertex is an active user ID and an edge represents a Bitcoin payment transferred between two user IDs. The bitcoin-au dataset consists of all transactions made from September 2010 to December 2013.   
    \item Online social networks [college-msg~\cite{panzarasa2009patterns}; Facebook-wall~\cite{viswanath2009evolution}]: Each vertex is a user who interacts with others through an online platform. The college-msg network, spanning over 193 days, originates from an online community for students at University of California, Irvine, and an edge ($v_i, v_j$) means user $v_i$ sent or received a private message to or from user $v_j$. The Facebook-wall dataset is derived from the Facebook New Orleans networks ranging from September 2004 to January 2009, where an edge ($v_i, v_j$) indicates user $v_i$ (or $v_j$) posted on user $v_j$'s (or $v_i$'s) wall.
    \item Short message correspondence [short-msg~\cite{wu2010evidence}]: Each vertex is a mobile phone user and an edge ($v_i, v_j$) represents user $v_i$ sent or received a short message (SM) to or from user $v_j$. The short-msg dataset spans over 338 days. 
\end{itemize}
\begin{table}[!htbp]
\caption{Summary statistics of datasets}
\centering
\begin{tabular}{l|c|c|c|c}
\hline 
Dataset & \# vertices & \# edges & \# periods & $(V_t, E_t)$ \\
\hline
email-Enron & 182 & 2097 & 30 & (150$\pm$ 31, 1,150$\pm$ 674) \\
\hline
bitcoin-au & 1288 & 6236 & 24 & (739 $\pm$ 390, 3,164 $\pm$ 2,024)\\
\hline
college-msg & 1899& 13,838 & 23 & (1,787 $\pm$ 93, 12,847 $\pm$ 923)\\
\hline
email-Eu-core & 893 & 12,556 & 51 & (823 $\pm$ 56, 8,542 $\pm$ 2,758)\\ 
\hline
short-msg & 44,090 & 52,222 & 41 & (30,209 $\pm$ 9,485, 30,952 $\pm$ 12,816)\\
\hline
Facebook-wall & 45,813 & 183,412 & 37 & (16,797 $\pm$ 11,819, 59,711 $\pm$ 50,379)\\
\hline
\end{tabular}
\label{table_dataset_stats}
\end{table}

More detailed information about our datasets is summarized in Table~\ref{table_dataset_stats}. The second and third column refers to the number of vertices and edges at the end of the observation period, respectively. From $(V_t, E_t)$, we observe that the college-msg dataset is a relatively slow-growing network while the Face-wall dataset is a fast-growing one. This is consistent with the values shown in the second column from Table~\ref{table_update_stats}, as less updates are needed for the college-msg dataset while more updates are performed for the Facebook-wall one. We also notice the corresponding values for the other four datasets are about 0.5. This is because for steadily growing networks, the probably that the TSN bound is larger than the average TSN bound from the previous window is 0.5. Thus, for networks with fairly steady growth, the fraction of periods at which CT is updated should be around 0.5.
\begin{table}[!htbp]
\caption{Summary statistics relative to the evolution of networks}
\centering
\begin{tabular}{l|c|c}
\hline 
Dataset & TSN bound & Fraction of periods at which CT is updated\\
\hline
email-Enron & 31 $\pm$ 16 & 0.5\\
\hline
bitcoin-au & 76 $\pm$ 27 & 0.5\\
\hline
college-msg & 75 $\pm$ 73 & 0.22\\
\hline
email-Eu-core & 106 $\pm$ 38 & 0.45\\ 
\hline
short-msg & 944 $\pm$ 493 & 0.56\\
\hline
Facebook-wall  & 2,981 $\pm$ 2,092 & 0.78\\
\hline
\end{tabular}
\label{table_update_stats}
\end{table}
\begin{figure}[!htbp]
\centering
\subfloat[]{\includegraphics[width=3.2in]{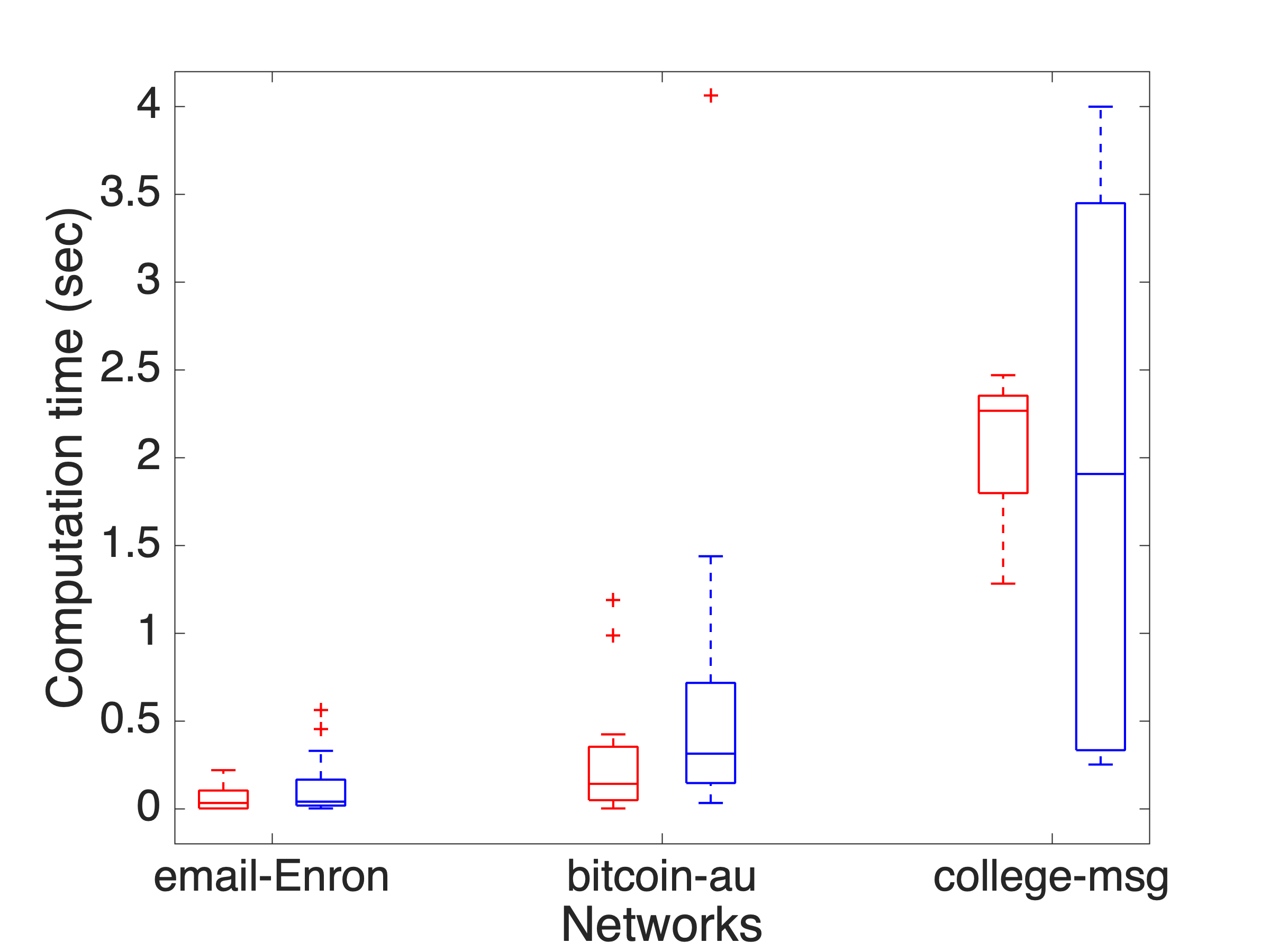}}
\subfloat[]{\includegraphics[width=3.2in]{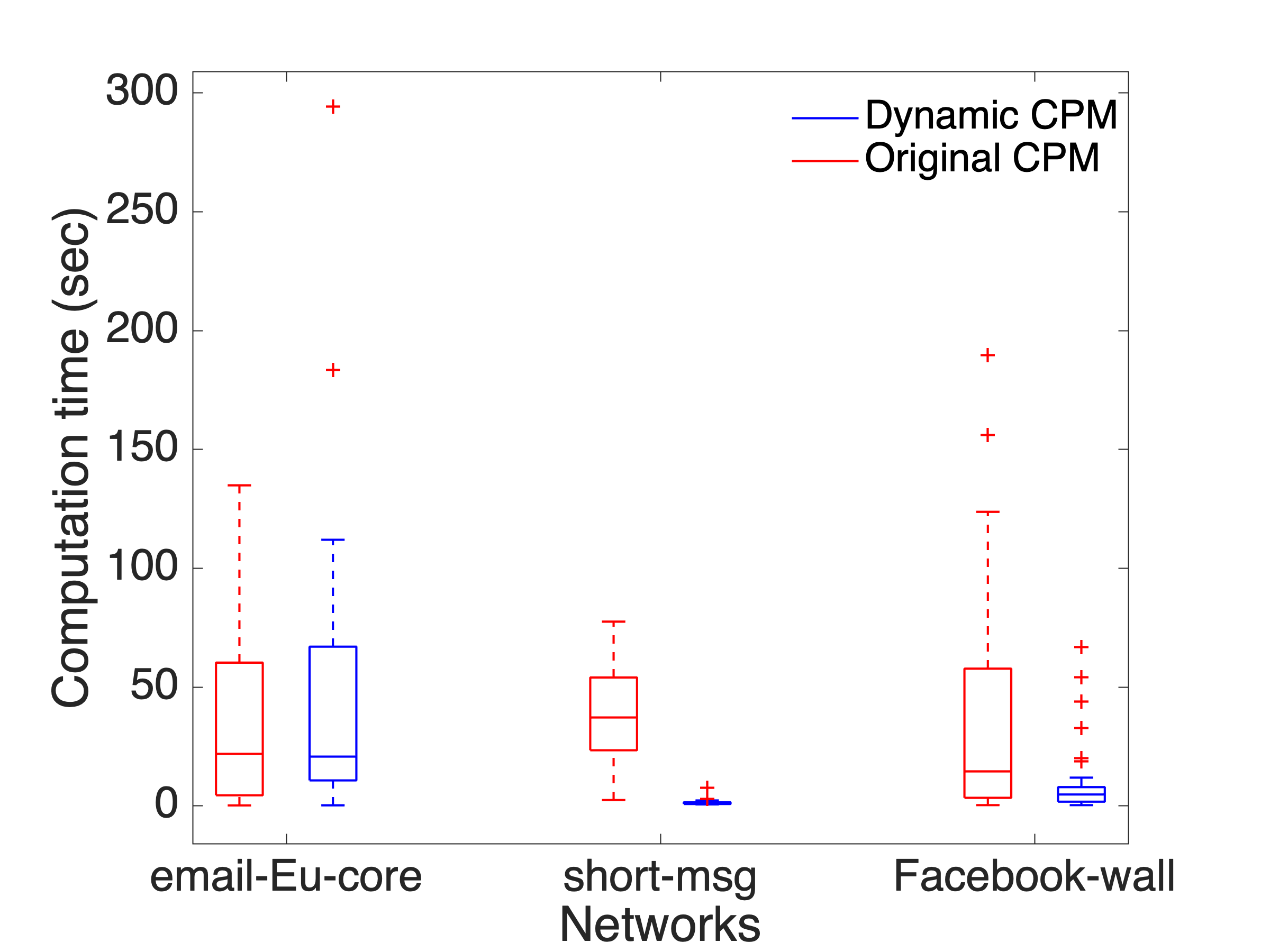}}
\caption{Comparison of computation time between the original CPM and our algorithm for six network datasets. The networks are divided into two groups based on their computation time for display purposes only.}
\label{fig_network_time}
\end{figure}
Figure~\ref{fig_network_time} compares the computation time, including building and updating community trees using the Dynamic CPM to the cumulative computation time for orders above 1 using original CPM as measured on a desktop with a 2.2 GHz Intel Xeon processor and 32 GB RAM. It can be seen that our algorithm reduces the computation time significantly on the two largest datasets, short-msg and Facebook-wall, while there is no significant computation time difference between the two methods for the other four datasets. 

This observation is further verified by the statistical test in Table~\ref{table_comp_time_stats_test}. We further evaluate three quantities that are often used to measure small world effect on these networks. As shown in Table~\ref{table_comp_time_stats_test}, the average clustering coefficients of the short-msg and Facebook-wall datasets are considerately smaller than other datasets, and their average shortest path lengths and effective diameters~\cite{leskovec2007graph} are noticeable larger than the others. All these values reflect that the communities in short-msg and Facebook-wall networks are less close-knit than the other networks over time. In fact, the short-msg and Facebook-wall networks can be classified as small world networks, as their average shortest path lengths scale with $\ln V_t$. Meanwhile, the other networks are regarded as ultra-small world since they have smaller average shortest path lengths that scale with $\ln \ln V_t$~\cite{barabasi2016network}. The corresponding community trees for small world networks tend to have more branches, especially at lower orders. The highest order of these trees are also much smaller than their counterparts for the ultra-small networks. This structural characteristics implies that the \Call{Link}{} and/or \Call{Cut}{} operations during the insertion and/or deletion of CG edges are more often performed between splay trees of small sizes, as opposed to the scenario for ultra-small networks where large splay trees are more frequently involved during these operations. This explains why our algorithm shows more strength for small world networks than ultra-small ones.
\begin{table}[!htbp]
\caption{Wilcoxon rank-sum test for computation time difference between the original CPM and our algorithm, and measures for the degree of clustering in networks}
\centering
\begin{tabular}{l|c|c|c|c}
\hline 
\multirow{2}{*}{Dataset} & \multirow{2}{*}{$p$-value} & Average clustering & Average shortest & Effective diameter \\
& & coefficient & path length & (90th percentile) \\
\hline
email-Enron & 0.39 &0.44 $\pm$ 0.08 & 2.49 $\pm$ 0.38 & 3.12 $\pm$ 0.56 \\
\hline
bitcoin-au & 0.16 & 0.34 $\pm$ 0.02 & 2.82 $\pm$ 0.15 & 3.40 $\pm$ 0.23\\
\hline
college-msg & 0.84 & 0.11 $\pm$ 0.002 & 3.05 $\pm$ 0.12 & 3.61 $\pm$ 0.15\\
\hline
email-Eu-core & 0.61 & 0.36 $\pm$ 0.03 & 2.89 $\pm$ 0.29 & 3.44 $\pm$ 0.46\\ 
\hline
short-msg & \bf 3.9e-9 & \bf 0.045 $\pm$ 0.011 & \bf 9.70 $\pm$ 1.66 & \bf 13.2 $\pm$ 2.25\\
\hline
Facebook-wall & \bf 0.02 & \bf 0.097 $\pm$ 0.025 & \bf 6.20 $\pm$ 1.44 & \bf 7.68 $\pm$ 2.05\\
\hline
\end{tabular}
\label{table_comp_time_stats_test}
\end{table}

\begin{figure}[!htbp]
\centering
\includegraphics[width=5.8in]{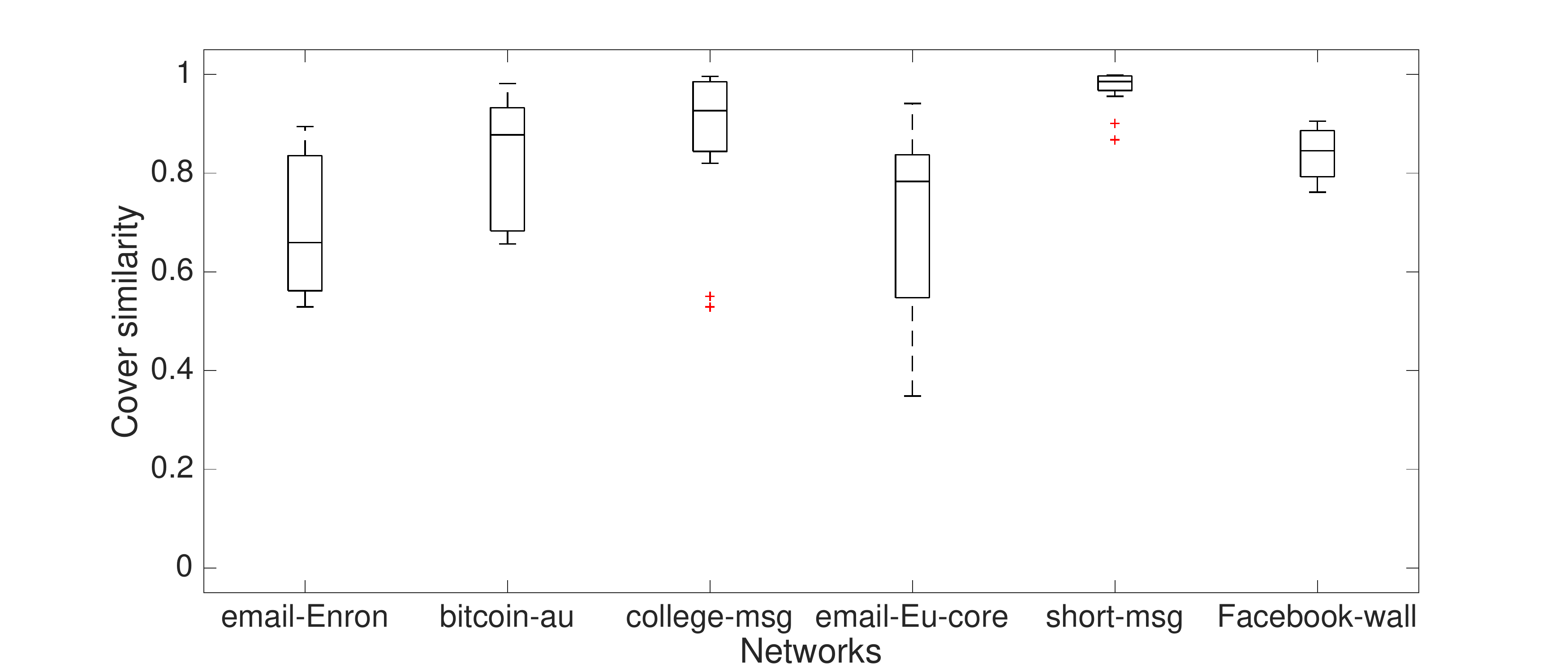}
\caption{Distribution of similarities measure by NMI between a set of 3-communities found in a network up to a particular time and the cover found by the previous update of the network for our datasets.}
\label{fig_network_acc}
\end{figure}
Figure~\ref{fig_network_acc} evaluate the similarities between the network cover\footnote{A set of communities is often referred to as a cover in the literature} from an unchanged community tree at order 3 and the real set of 3-communities by $t_i$ for all the datasets. The evaluation metric is chosen to be normalized mutual information (NMI), which is a measure of similarities between the partitions based on information theory. This metric is proven to be reliable for comparing the real communities and the found ones when evaluating community detection algorithms~\cite{danon2005comparing}. Here we adopt a widely-used version, proposed by Lancichinetti et al.~\cite{lancichinetti2009detecting}, for comparative analysis of overlapping communities. It can be seen that the mean values of the cover similarity for small-world networks are roughly above 0.85, while those vary from around 0.65 to 0.9 among ultra-small world networks.
\begin{figure}[!bp]
\centering
\includegraphics[width=\columnwidth]{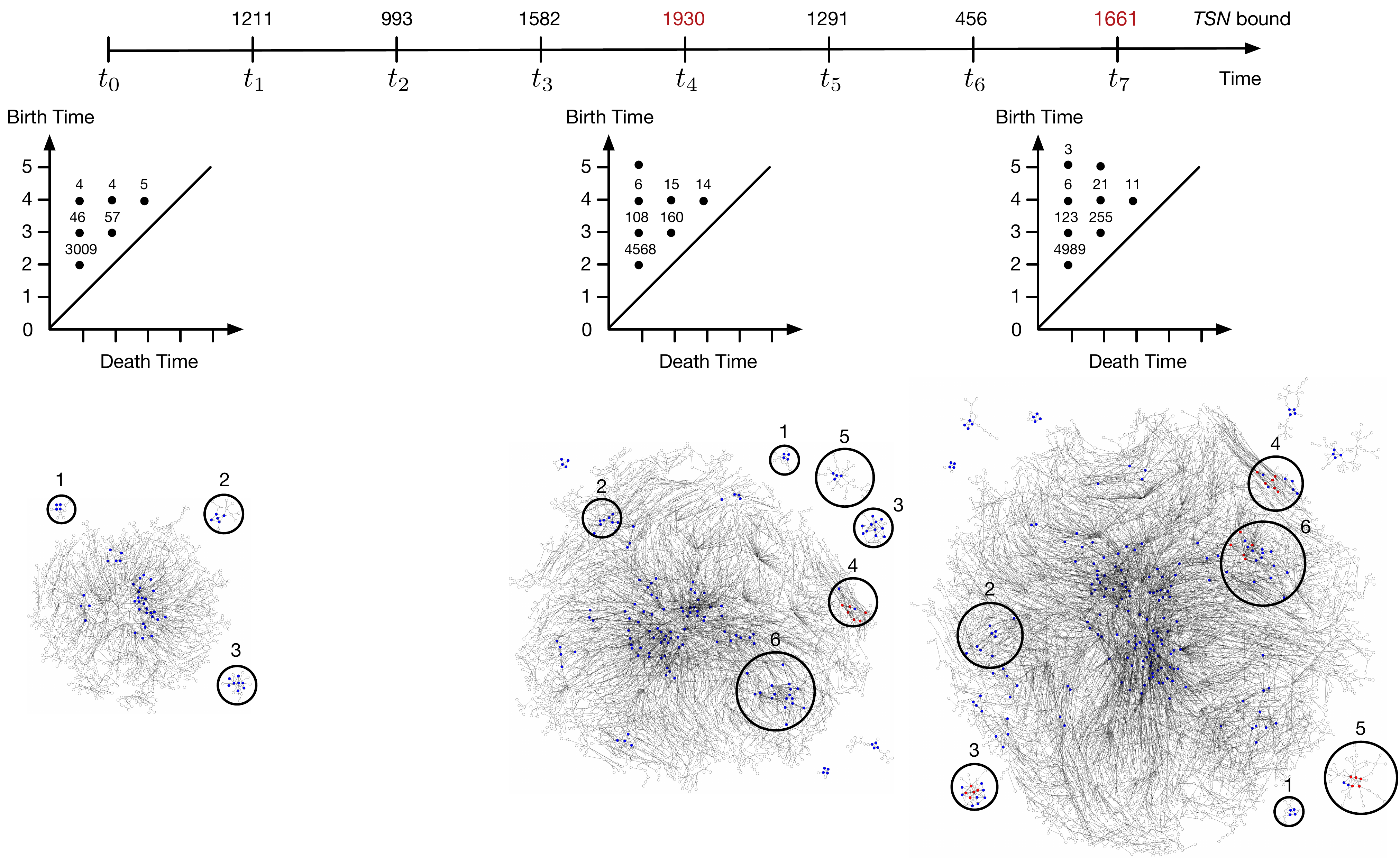}
\caption{Network evolution for the short-msg dataset. The PDs correspond to the community trees at $t_0$, $t_4$, $t_7$, respectively. The number attached above to a point represents the multiplicity of this point in the PDs. The graphs shown at the bottom are a collection of CCs (subgraphs) where the set of 4-communities (blue) and 5-communities (red) belong to.}
\label{fig_sms_evolution}
\end{figure}

In Figure~\ref{fig_sms_evolution}, the short-msg dataset serves an example to demonstrate the evolutionary network analysis using our algorithm. The community tree is initially built at $t_0$ and updated at $t_4$ and $t_7$ base on the rule in our algorithm. The corresponding PDs indicate that the community tree is featured with a large number of lower order branches throughout the duration. The generated community tree by our algorithm is able to track communities over time and identify their appearance and disappearance by their labels. In this example, the highest order of $\mathcal{T}_0$ is 4 and there are four CCs at $t_0$. From $t_0$ to $t_4$, a newborn 5-community CC$_4$ appears, and it is not evolved from any of the 4-communities at $t_0$. CC$_1$ and CC$_3$ stay isolated from others, but CC$_2$ has been merged to the central CC at $t_4$. From $t_4$ to $t_7$, three more 5-communities are born, where one of them is CC$_3$ and the other two, CC$_5$ and CC$_6$ are evolved from newly 4-communities at $t_4$. Interestingly, CC$_1$ still remains disconnected to any other CCs. 

\section{Discussion and Future Work}
\label{sec: network_TDA_discussion}
In this paper, we develop an algorithm called the Dynamic CPM that can efficiently detect communities in large-scale dynamic networks. The experiments show that our algorithm reduces the computation time significantly compared to the original CPM, especially for small world networks. The network cover of 3-communities from unchanged community trees also maintain relatively high similarities to the real covers. Furthermore, since the community tree encodes the evolutionary information of communities, our algorithm can naturally track similar communities over time, which often helps to determine fundamental structures of dynamic networks. In addition, the concise tree structure also records when communities appear, disappear, split or merge, which allows us to identify the occurrence of critical events.

The framework in this paper is expected to lay the basis for a TDA route to the study of multilayer networks, particularly in fragility analysis.
Multilayer networks exhibit a more realistic fashion to characterize a wide variety of real-world networks~\cite{jeub2017local}, where intra-layer edges encode different or related types of interactions, and dynamical processes traverse along both intra-layer and inter-layer edges. 
In fact, a temporal network is a special type of multilayer network where each time instant is mapped into a different layer. Since 
random failures or targeted attacks tend to cause cascading effects (the failure of one node will recursively provoke the failure of connected nodes) on the network functionality and the magnitude of such effects is directly related to the topology of the network, investigating community-based topological structures and the resilience of multilayer networks will help to improve infrastructural design in relevant applications so as to make them more robust to critical failure modes.

\bibliographystyle{unsrt}  
\bibliography{references}  

\clearpage
\begin{center}
\textbf{\large Supplemental Materials: Efficient Community Detection in Large-Scale \\ Dynamic Networks Using Topological Data Analysis}
\end{center}
\setcounter{section}{0}
\renewcommand{\thesection}{S\arabic{section}}
\setcounter{equation}{0}
\setcounter{figure}{0}
\setcounter{table}{0}
\makeatletter
\renewcommand{\theequation}{S\arabic{equation}}
\renewcommand{\thefigure}{S\arabic{figure}}
\renewcommand{\thetable}{S\arabic{table}}

\section{Design of Data Structure}
\label{sec:data-structure}

\begin{figure}[!bp]
\center
\includegraphics[width=0.6\columnwidth]{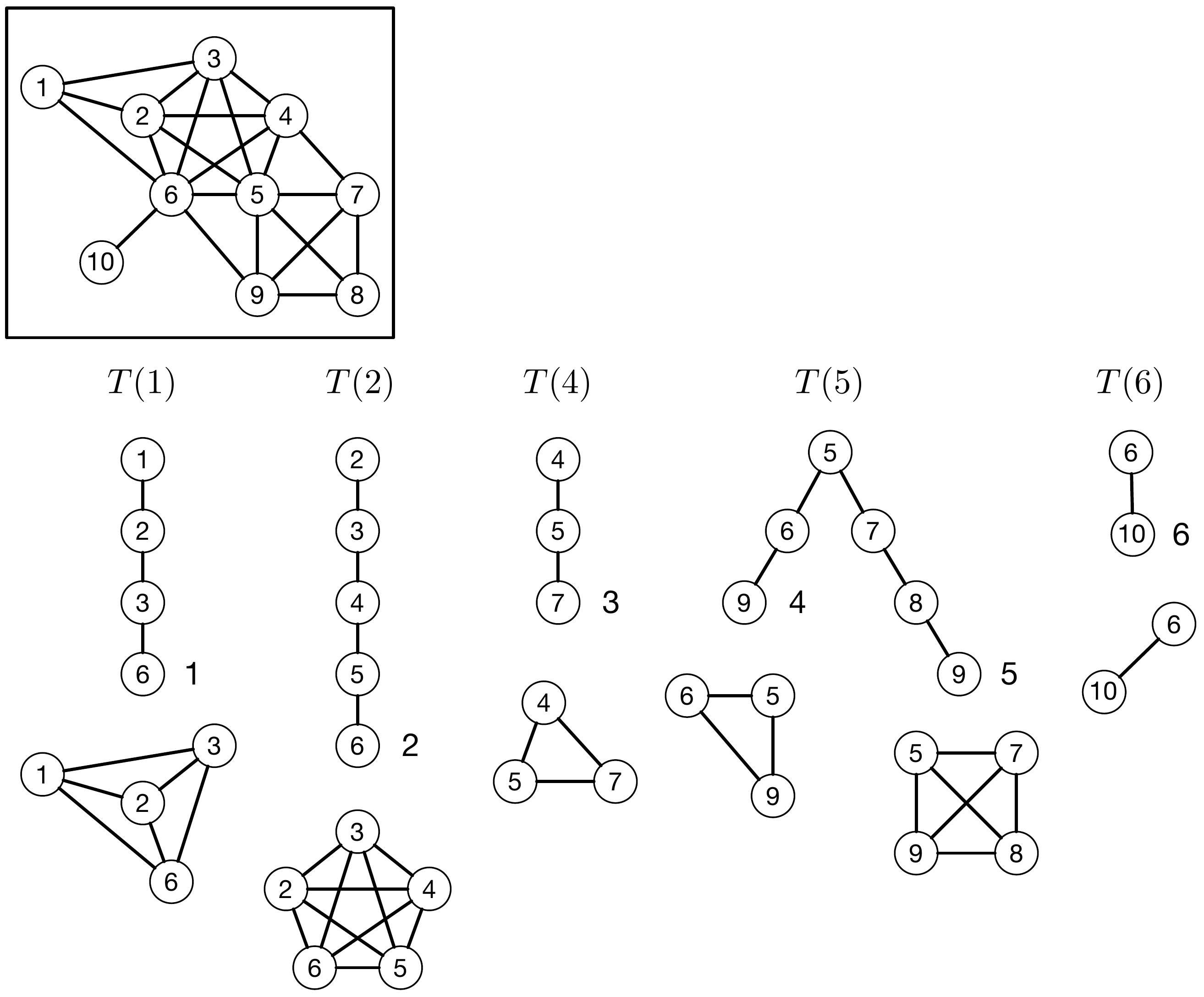}
\caption{Tries for the network shown in Figure~\ref{fig::CG}. For $i=1,2,\ldots,6$, each root-to-leaf path in $T(i)$ represents an MC whose lowest labeled vertex is $i$. For example, there are two MCs containing vertex $5$, where all the other vertices have labels greater than $5$. One tree path is an MC of size $3$ containing vertices $5,6,9$. The other path is an MC of size $4$ comprising vertices $5,7,8,9$. Each MC's ID is stored at the leaf node of its trie.} 
\label{fig::radixTr}
\end{figure}

We assign each vertex $v$ with a unique ID, denoted by $v.id$, and sort each MC in an increasing order by the vertex ID. Let $T(v)$ be a \textbf{trie} that stores the set of MCs of $G$ starting with root $v$. That is, for any MC $M \in T(v)$, $v.id=\min \{u.id: u \in M\}$.
 Each MC, represented by a root-to-leaf path, is also affiliated with a unique ID in an increasing order and stored at the leaf node. Figure~\ref{fig::radixTr} provides an example of tries rooted at several vertices of a graph. This form of tree is a space-optimized data structure. It only stores the vertices in the shared root-to-node subpath between the MCs once, and, therefore, uses less space than a hash table with the nodes as keys and MCs as values. It does not have to be binary. It also supports search, insertion, and deletion operations in time of the order of the number of elements $k$ in the operation set. Specifically, the functions used in the following algorithms relevant to a trie include: 

\begin{itemize}[nosep]
\item $\Call{Initialize-trie}{v, m}$: Initialize a trie $T(v)$ with ID $m$
\item $\Call{Trie-add}{T(v), M, m}$: Add the MC $M$ with ID $m$ to $T(v)$ 
\item $\Call{Trie-remove}{T(v), M}$: Remove the MC $M$ from $T(v)$
\item $\Call{Trie-get-id}{T(v), M}$: Obtain the ID of $M$ from $T(v)$
\end{itemize}

To support efficient querying, we record the following information for each node $x$ in $T(v)$:
\begin{itemize}[nosep]
\item {\it key}: The ID of vertex $u$ that node $x$ represents.
\item {\it value}: The value is initiated as -1. If $x$ is a leaf node, then it stores the ID of the MC.
\item {\it parent}: The parent node of $x$.
\item {\it child}: A hash table that stores the children nodes of $x$.
\item {\it next}: The next node of $x$ in the linked list stored in {\it nodeList}.
\end{itemize}

We also define the following data fields of $T(v)$: 

\begin{itemize}[nosep]
\item {\it root}: The root node of $T(v)$.
\item {\it nodeList}: A hash table that maps the ID of a certain vertex $u$ to a linked list of nodes in $T(v)$ that represent $u$ in each entry; this list is required as the same vertex $x$ may appear multiple times in different branches of $T(v)$~\cite{distributed_Xu_2016}.
\end{itemize}

To manage the collections of $T(v)$ for all the vertices $v \in V(G)$, we create a hashtable $\mathscr{T}$ with each $v.id$ as the key. The three operations on the hash table are listed below:
\begin{itemize}[nosep]
\item $\Call{HT-get}{\mathscr{T}, v.id}$: Obtain $T(v)$ from $\mathscr{T}$ through $v.id$
\item $\Call{HT-add}{\mathscr{T}, (v.id, T(v))}$: Add $(v.id, T(v))$ to $\mathscr{T}$
\item $\Call{HT-delete}{\mathscr{T}, v.id}$: Delete $(v.id, T(v))$ from $\mathscr{T}$ through $v.id$
\item $\Call{HT-iskey}{\mathscr{T}, v.id}$: True if $v.id$ is a key in $\mathscr{T}$ 
\end{itemize} 

The set of MCs of $G$ is computed using the pivoting-based depth-first search algorithm presented in~\cite{worst_Tomita_2006}, which has been shown to be worst-case optimal with a runtime of $O(3^{n/3})$ for an $n$-vertex graph. Table~\ref{table_symbols} summarizes the symbols used in this Appendix. 

\begin{table}[!hb]
\begin{center}
\caption{Summary of symbols and descriptions}
\begin{tabular}{c|l}
\hline
Symbol & Description \\
\hline
$M$ & Maximal clique (MC)\\
\hline
$E_{ab}$ & ET tree node formed of CG nodes $a$ and $b$ \\
\hline
$E_{M}$ & ET tree node corresponding to $M$ \\
\hline
$E_{R}$ & Root node of an ET tree \\
\hline 
$\mathcal{G}_i$ & Weighted CG at level $i$ \\
\hline
$L(v)$ & Set of the vertices adjacent to $v$ whose IDs are smaller than $v$'s ID \\
\hline
$m$ & ID of $M$ \\
\hline
$\mathcal{M}(G)$ & Set of MCs in the graph $G$ \\
\hline
$\omega(G)$ & Maximum size of MCs in $G$\\
\hline
$\mathsf{M}$ & MC object based on $M$\\
\hline
$\mathscr{M}$ & Hash table with the (key, value) pair as $(m, \mathsf{M})$ for each entry \\
\hline
$n_M$ & Array of CG nodes with each element representing an $M$ in $\mathcal{G}_i$ \\
\hline
$P$ & PD represented by an array of birth and death time pairs of communities $\mathcal{C}$\\ 
\hline
$T(v)$ & Trie that stores the set of MCs starting with root $v$\\
\hline
$\mathscr{T}$ & Hash table with the (key, value) pair to be $(v.id, T(v))$, $v \in V$ for each entry \\
\hline
$\mathcal{T}$ & Community tree of $G$ represented by an array of size $\omega(G)$\\
\hline
\end{tabular}
\label{table_symbols}
\end{center}
\end{table}

\section{Initial Community Tree Construction Algorithm}
\label{sec::initial-construction}

It has been explained in Section~\ref{sec: CG_and_ETT} that building a network $G_0$'s community tree amounts to generating its corresponding spanning forests and ET trees at each level. We divide the generation process into two steps, as is given in Algorithm~5 and illustrated in Figure~\ref{fig::buildCT}. Note that all the operations related to spanning trees include the operations involved with ET trees if necessary.

Specifically, the first step is to initialize the sequence of the CGs (lines 3-15 in Algorithm~5), i.e., place each existing MC of size $|M| \geq 2$ as a CG node to each associated CG. Thus, for each $M \in \mathcal{M}(G_0)$ with $|M| \geq 2$, we initialize a MC object $\mathsf{M}$ given its ID $m$ and its size $|M|$ first, and $\mathsf{M}$ becomes the initial representative MC of the single-MC component. As aforementioned in Section~\ref{sec: CG_and_ETT}, the birth and death time of a CC is recorded as those of its representative MC. Hence, each $\mathsf{M}$ in Algorithm~\ref{alg::initialize-MC} possesses the following data fields: {\it id}, {\it visited}, {\it birthT}, {\it deathT}, {\it child}, and holds a pointer $\mathsf{M}.pCG$ to an array of CG nodes. 
\begin{algorithm}[!b]
\caption{Build the initial $\mathcal{T}$ for a given unweighted, undirected graph $G_0$} 
\label{alg::buildCT}
\begin{algorithmic}[1]
\Procedure{Build-CT}{$G_0$} 
\State $\mathscr{T} \gets \emptyset$, $\mathscr{M} \gets \emptyset$, $\mathcal{T}' \gets \emptyset$, $m_0 \gets 0$
\State $\mathcal{M}(G_0) \gets \Call{List-MCs}{G_0}$ \Comment Algorithm in~\cite{worst_Tomita_2006}
\State $\omega(G_0) \gets \max_{M \in \mathcal{M}(G_0)}|M|$ 
\State Initialize an empty array of size $\omega(G_0)$ as $\mathcal{T}$
\State $\mathscr{T}, \mathscr{M}, \mathcal{T}, m \gets \Call{Add-MCs}{G_0\mathcal{M}(G_0), \mathscr{T}, \mathscr{M}, \mathcal{T}, \mathcal{T}', m_0}$ \Comment Algorithm~5

\Comment Extract the birth and death time for nodes in the community tree 
\State $\mathscr{M}, P \gets \Call{Generate-PD}{\mathscr{M}, \mathcal{T}}$ \Comment Algorithm~11

\State \Return $\mathscr{T}, \mathscr{M}, \mathcal{T}, P, m$
\EndProcedure 
\end{algorithmic}
\end{algorithm}
In our pseudo-code, any variable that represents an array or object is treated as a {\it pointer} to the data representing the array or object. 
For each CG node $\mathsf{M}.pCG[i]$, $i=|M|-1, \ldots, 1$, its attributes include {\it id}, {\it size} and {\it adjTrEdges} that represent the MC's ID and size, as well as the adjacency lists of the tree edges, respectively. Furthermore, each CG node also has a pointer to its ET node in the corresponding ET tree. For each ET node $E_{ab}$, in turn, formed of two CG nodes $a$ and $b$, the node value affiliated with $E_{ab}$ is:
\begin{figure}[!t]
\centering				
\includegraphics[width=0.8\columnwidth]{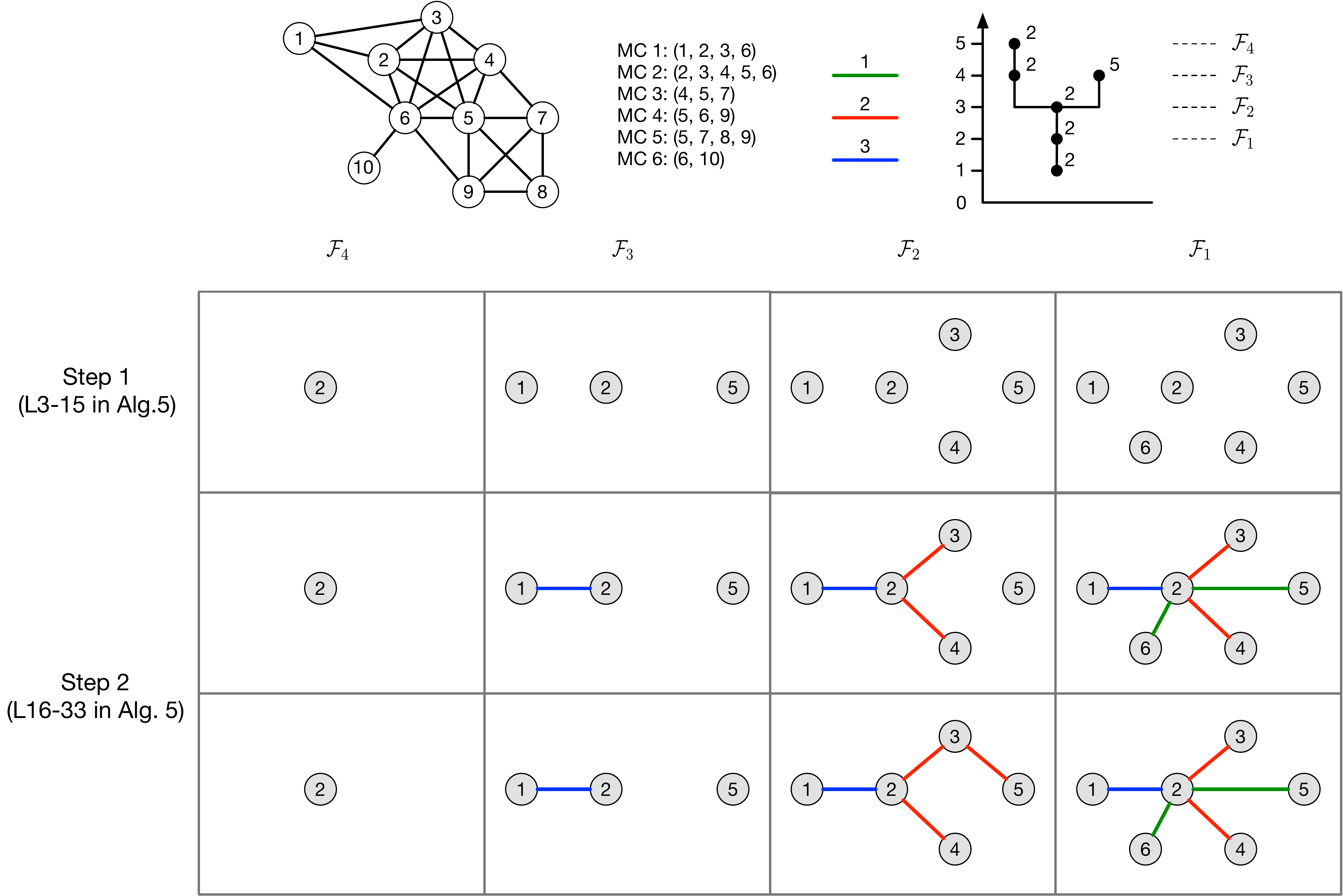}
\caption{An example of building a community tree. To construct a community tree, each existing MC of size $|M| \geq 2$ is first laid as a single-node spanning tree in $\mathcal{F}_{|M|-1}, \ldots, \mathcal{F}_1$, respectively (top row). Starting with CG node 2 arbitrarily, we then build connections between node 2 and its neighbors, i.e., nodes 1, 3, 4, 5 and 6. For each unvisited neighbor, we insert a tree edge with weight $w_{e_{\mathcal{G}}}$ in $\mathcal{F}_i$, $i \leq w_{e_{\mathcal{G}}}$ between the two nodes if they are from different spanning trees (middle row). Node 2 is marked as a visited one after all the connections. We repeat this process for node 3, and thus, add a tree edge to node 5 in $\mathcal{F}_2$ (bottom row). No more connections are made after we go through the other nodes.} 
\label{fig::buildCT}
\end{figure}

\begin{algorithm}[tbp]
\caption{Add each MC in $\mathcal{M}^{new}$ with its corresponding CG nodes and ET nodes}
\begin{algorithmic}[1]
\Procedure{Add-MCs}{$G, \mathcal{M}^{new}, \mathscr{T}, \mathscr{M}, \mathcal{T}, \mathcal{T}', m_0$}
\State $m \gets m_0$

\Comment Initialize the CGs at each order
\ForEach {$M \in \mathcal{M}^{new}$} 
\State $m \gets m+1$
\State $v^* \gets \argmin_{v \in M} v.id$
\State $T(v^*) \gets \Call{HT-get}{\mathscr{T}, v^*.id}$
\If {$T(v^*)=\textrm{NULL}$}
\State $T(v^*) \gets \Call{Initialize-trie}{v^*, m}$
\State $\mathscr{T} \gets \Call{HT-add}{\mathscr{T}, (v^*.id, T(v^*))}$
\EndIf
\State $T(v^*) \gets \Call{Trie-add}{T(v^*), M, m}$

\If {$|M|\geq 2$}
\State $\mathsf{M} \gets \Call{Initialize-MC}{|M|, m}$ \Comment Algorithm~\ref{alg::initialize-MC}
\State $\mathscr{M} \gets \Call{HT-add}{\mathscr{M}, (m, \mathsf{M})}$
\For {$i=|M|, \ldots, 2$}
\State $\mathcal{T}[i] \gets \Call{HT-add}{\mathcal{T}[i], (m, \mathsf{M}.pCG[i-1].pET)}$
\EndFor
\EndIf 
\EndFor

\Comment{Generate spanning forests by connecting the nodes based on their edge weights}

\ForEach {$M \in \mathcal{M}^{new}$} 
\State $v^* \gets \argmin_{v \in M} v.id$
\State $T(v^*) \gets \Call{HT-get}{\mathscr{T}, v^*.id}$
\State $m \gets \Call{Trie-get-id}{T(v^*), M}$
\State $\mathsf{M} \gets \Call{HT-get}{\mathscr{M}, m}$
\State $\mathcal{N}(M) \gets \Call{Get-neighbor-MCs}{G, M, m, \mathscr{T}}$ 
\Comment Algorithm~\ref{alg::get-neighbor-MCs} 
\If {$\mathcal{N}(M) \neq \emptyset$}
\ForEach {$(m',M') \in \mathcal{N}(M)$} 
\State $\mathsf{M'} \gets \Call{HT-get}{\mathscr{M}, m'}$

\If {$\mathsf{M'}.visited=\textrm{TRUE}$ \textbf{and} $m'> m_0$}
\Continue
\Else
\If {$\min(|M|,|M'|)=2$}
\State $\mathcal{T} \gets \Call{Insert-CG-edges}{\mathsf{M}.pCG, \mathsf{M'}.pCG, \mathcal{T}, \mathcal{T}', 1}$ \Comment Algorithm~\ref{alg::insert-edges}
\Else
\State $w_{e_{\mathcal{G}}} \gets \Call{Intersect}{M, M'}$
\State $\mathcal{T} \gets \Call{Insert-CG-edges}{\mathsf{M}.pCG, \mathsf{M'}.pCG, \mathcal{T}, \mathcal{T}', w_{e_{\mathcal{G}}}}$
\EndIf
\EndIf
\EndFor
\EndIf
\State $\mathsf{M}.visited=\textrm{TRUE}$
\EndFor

\State \Return $\mathscr{T}, \mathscr{M}, \mathcal{T}, m$
\EndProcedure 
\end{algorithmic}
\label{alg::add-MCs}
\end{algorithm}
\begin{itemize}[nosep]
\item {\it dRepID}: Defined by Equation~(\ref{eq::delta-w}), where $w(x)$ is the ID of the representative ET node of the ET tree that $E_{ab}$ belongs to if $E_{ab}$ is a loop node, and $w(x)$ is assigned to be 0 if $E_{ab}$ is a nonloop node. Thus, for a single-node ET tree, $E_{ab}.dRepID=a.id$ if $E_{ab}$ is a loop node, while $E_{ab}.dRepID = 0$ if $E_{ab}$ is a nonloop one.
\end{itemize} 
\begin{algorithm}[tbp]
\caption{Initialize an MC $\mathsf{M}$ given its ID $m$ and its size $|M|$}
\begin{algorithmic}[1]

\Procedure{Initialize-MC}{$|M|, m$}
\State $\mathsf{M}.id \gets m$
\State $\mathsf{M}.visited \gets \textrm{FALSE}$
\State $\mathsf{M}.birthT \gets |M|$ 
\State $\mathsf{M}.deathT \gets -1$
\State $\mathsf{M}.child \gets \emptyset$
\State $\mathsf{M}.pCG \gets \Call{Set-CG-node}{|M|,m}$
\State \Return $\mathsf{M}$
\EndProcedure \\

\Function{Set-CG-node}{$|M|, m$} \Comment Initialize a CG node in $\mathcal{G}_{|M|-1}, \ldots, \mathcal{G}_1$, respectively
\State Initialize an array of size $|M|-1$ as $n_M$
\For {$i = 1, \ldots, |M|-1$}
\State $n_M[i].id \gets m$ \Comment $n_M[i]$ represents the CG node in $\mathcal{G}_i$
\State $n_M[i].size \gets |M|$
\State $n_M[i].adjTrEdges \gets \emptyset$ \Comment Adjacency list of tree edges
\State $n_M[i].pET \gets \textrm{NULL}$
\EndFor\For {$i = 1, \ldots, |M|-1$}
\State $n_M[i].pET \gets \Call{Set-ET-node}{n_M[i], n_M[i]}$ 
\Comment Point to its ET node
\EndFor 
\State \Return $n_M$
\EndFunction \\

\Function{Set-ET-node}{$a, b$} \Comment Create an ET node with CG nodes $a$ and $b$
\State $E_{ab}.CG_1 \gets a$, $E_{ab}.CG_2 \gets b$
\State $E_{ab}.parent \gets \textrm{NULL}$,
$E_{ab}.right \gets \textrm{NULL}$, 
$E_{ab}.left \gets \textrm{NULL}$
\If {$a.id=b.id$} \Comment If the ET node corresponds to a node in CG
\State $E_{ab}.dRepID \gets a.id$
\Else \Comment If the ET node corresponds to a tree edge in CG
\State $E_{ab}.dRepID \gets 0$
\EndIf
\State \Return $E_{ab}$ 
\EndFunction
\end{algorithmic}
\label{alg::initialize-MC}
\end{algorithm}
\begin{algorithm}[!tb]
\caption{Find the neighboring MCs given a MC $M$ and its ID $m$}
\begin{algorithmic}[1]
\Procedure{Get-neighbor-MCs}{$G, M, m, \mathscr{T}$}

\State $\{v_1,\ldots,v_{|M|}\} \gets M$, $\mathcal{N}(M) \gets \emptyset$

\Comment Output all the MCs except $M$ in $T(u)$ for $u \in M$
\For {$j=1, \ldots, |M|$} 
\State $T(v_j) \gets \Call{HT-get}{\mathscr{T}, v_j.id}$
\State $\mathcal{N}(v_j) \gets \Call{Output-MC}{T(v_j).root, T(v_j)}$ \Comment{Algorithm~\ref{alg::output_mc}}
\State $\mathcal{N}(M) \gets \mathcal{N}(M) \cup \mathcal{N}(v_j)$
\EndFor 
\State $\mathcal{N}(M) \gets \Call{HT-delete}{\mathcal{N}(M), m}$

\Comment Output all the MCs in $T(u)$ for $u \in \tilde{L}(v_j)$ that contains $v_j$, $j=1, \ldots, |M|$

\For {$j=1, \ldots, |M|$} 
\State $L(v_j) \gets \Call{Get-lower-neighbors}{G, v_j.id}$
\State $\tilde{L}(v_j) \gets L(v_j)\backslash\{v_i: v_i \in M, v_i.id < v_j.id\}$ 
\ForEach {$u \in \tilde{L}(v_j)$}
\State $T(u) \gets \Call{HT-get}{\mathscr{T}, u.id}$
\State $x \gets \Call{HT-get}{T(u).nodeList, v_j.id}$
\While{$x \neq$ NULL}
\State $\mathcal{N}(x) \gets \Call{Output-MC}{x, T(u)}$ 
\State $\mathcal{N}(M) \gets \mathcal{N}(M) \cup \mathcal{N}(x)$
\State $x \gets x.next$
\EndWhile
\EndFor
\EndFor
\State \Return $\mathcal{N}(M)$  
\EndProcedure 

\end{algorithmic}
\label{alg::get-neighbor-MCs}
\end{algorithm}
\begin{algorithm}[!tb]
\caption{Output all root-to-leaf paths that contain a node $x$ in $T(v)$}
\begin{algorithmic}[1]
\Function{Output-MC}{$x, T(v)$} 
\State $M \gets \emptyset$, $\mathcal{N}(x) \gets \emptyset$ \Comment{$\mathcal{N}(x)$ is defined as a global variable}
\State $p \gets x$
\While{$p \neq$ NULL}
\State $M \gets M \cup \{p.key\}$
\State $p \gets p.parent$
\EndWhile
\State $M \gets \Call{Reverse}{M}$ \Comment{Reverse the array of $M$}
\State $\Call{Search-down}{M, x, T(v)}$
\State \Return $\mathcal{N}(x)$ 
\EndFunction  \\

\Function{Search-down}{$M, x, T(v)$}
\If {$x.child \neq \emptyset$}
\State $\mathcal{N}(x) \gets \mathcal{N}(x) \cup \{(x.value, M)\}$
\State \Return 
\Else
\ForEach {$(y.key, y) \in x.child$}
\State $M \gets M \cup \{y.key\}$
\State \Call{Search-down}{$M, y, T(v)$}
\State $M \gets \Call{Delete-last}{M}$ \Comment{Delete the last element in $M$}
\EndFor
\EndIf
\EndFunction

\end{algorithmic}
\label{alg::output_mc}
\end{algorithm}
Apart from these node values, $E_{ab}$ also stores three pointers to its parent, right and left child. To gain direct access to each $\mathsf{M}$, we create a hash table $\mathscr{M}$ in which the (key, value) pair for each entry is $(m, \mathsf{M})$. For a similar reason, we let $\mathcal{T}$ be an array of size $\omega(G_0)$ representing the community tree in which each element $\mathcal{T}[i]$ is a hash table with entries
$(m, \mathsf{M}.pCG[i-1].pET)$ for each representative MC $\mathsf{M}$ at order $i$ for $i \geq 2$ (lines 14-15 in Algorithm~5). $\mathcal{T}[i]$ gets updated after a tree edge is added in $\mathcal{F}_{i-1}$. 

The second step is to generate each $\mathcal{F}_i$ by connecting CG nodes based on their edge weights $w_{e_\mathcal{G}}$ (lines 16-33 in Algorithm~5). To avoid the expensive intersection tests against every other MC for a given $M \in \mathcal{M}(G_0)$ with $|M| \geq 2$, we first identify a set that consists of pairs of a neighboring MC $M'$ and its ID $m'$, denoted by $\mathcal{N}(M)$  (Algorithm~\ref{alg::get-neighbor-MCs}). Here, a `neighboring' MC of $M$ is an MC that shares at least one vertex with $M$.
We insert a tree edge between $\mathsf{M}.pCG[w_{e_\mathcal{G}}]$ and $\mathsf{M'}.pCG[w_{e_\mathcal{G}}]$ in $\mathcal{F}_{w_{e_\mathcal{G}}}$ as long as $\mathsf{M}.pCG[w_{e_\mathcal{G}}]$ and $\mathsf{M'}.pCG[w_{e_\mathcal{G}}]$ are not reachable, and repeat the insertion procedure in $\mathcal{F}_i, i <w_{e_\mathcal{G}}$, until $\mathsf{M}.pCG[i]$ and $\mathsf{M'}.pCG[i]$ are found in the same spanning tree.
Since the maximum possible number of vertices shared by $M$ and $M'$ is $\min (|M|,|M'|)-1$, we have $1 \leq w_{e_\mathcal{G}} \leq \min (|M|,|M'|)-1$. Therefore, the intersection tests are only performed between $M$ and its neighboring MCs when $\min (|M|,|M'|)>2$. For the special case when $\min (|M|,|M'|)=2$, $w_{e_\mathcal{G}}$ is 1; hence, no intersection test is required.
All the MCs are initially {\it unvisited}, and one is marked as {\it visited} once it finishes connections with all its unvisited neighboring MCs to avert duplicate calculations.

Finally, we record the death time of the representative MCs from the updated $\mathcal{T}$ and the corresponding PD is derived through a tree traversal (Algorithm~11). 

\begin{algorithm}[!tb]
\caption{Add an edge between $n_{M_1}[i]$ and $n_{M_2}[i]$ with weight $w_{e_{\mathcal{G}}}$ in $\mathcal{F}_i$, $i=w_{e_{\mathcal{G}}}, \ldots, 1$}
\begin{algorithmic}[1]
\Procedure{Insert-CG-edges}{$n_{M_1}, n_{M_2}, \mathcal{T}, \mathcal{T}', w_{e_{\mathcal{G}}}$}
\State $e \gets \Call{Set-CG-edge}{n_{M_1}, n_{M_2}, w_{e_{\mathcal{G}}}}$ 

\For {$i=w_{e_{\mathcal{G}}}, \ldots, 1$} 
\If {\Call{ETT-connected}{$n_{M_1}[i].pET, n_{M_2}[i].pET$} = \textrm{TRUE}}
\Break
\Else
\State $\Call{ETT-splay}{n_{M_1}[i].pET}$
\State $\Call{ETT-splay}{n_{M_2}[i].pET}$
\State $\mathcal{T} \gets \Call{Update-Rep}{n_{M_1}[i].pET, n_{M_2}[i].pET, \mathcal{T}, \mathcal{T}', i}$ \Comment Algorithm~\ref{alg::update-rep}
\State \Call{Add-tree-edge}{$n_{M_1}, n_{M_2}, e_{\mathcal{G}}, i$}
\EndIf
\EndFor
\State \Return $\mathcal{T}$
\EndProcedure \\

\Function{Set-CG-edge}{$n_{M_1}, n_{M_2}, w_{e_{\mathcal{G}}}$}

\Comment Initialize an edge with $w_{e_{\mathcal{G}}}$ inserted between CG nodes representing ${M_1}$ and ${M_2}$
\State $e_{\mathcal{G}}.CG_1 \gets n_{M_1}$, $e_{\mathcal{G}}.CG_2 \gets n_{M_2}$ 
\For {$i = 1, \ldots, w_{e_{\mathcal{G}}},$}
\State $e_{\mathcal{G}}.pET_1[i] \gets \textrm{NULL}$, $e_{\mathcal{G}}.pET_2[i] \gets \textrm{NULL}$
\EndFor
\State \Return $e_{\mathcal{G}}$
\EndFunction \\

\Function{Add-tree-edge}{$n_{M_1}, n_{M_2}, e_{\mathcal{G}}, i$}
\State $n_{M_1}[i].adjTEdges \gets \Call{Push-back}{n_{M_1}[i].adjTEdges, e_{\mathcal{G}}}$
\State $n_{M_2}[i].adjTEdges \gets \Call{Push-back}{n_{M_2}[i].adjTEdges, e_{\mathcal{G}}}$
\State $e_{\mathcal{G}}.pET_1[i] \gets \Call{Set-ET-node}{n_{M_1}[i], n_{M_2}[i]}$ \Comment Store the pointers in the edge
\State $e_{\mathcal{G}}.pET_2[i] \gets \Call{Set-ET-node}{n_{M_2}[i], n_{M_1}[i]}$ 
\State $\Call{ETT-link}{n_{M_1}[i].pET, n_{M_2}[i].pET}$
\EndFunction

\end{algorithmic}
\label{alg::insert-edges}
\end{algorithm}

\begin{algorithm}[!tb]
\caption{Update $\mathcal{T}$ before a tree edge insertion}
\begin{algorithmic}[1]
\Procedure{Update-Rep}{$E_{R_1}, E_{R_2}, \mathcal{T}, \mathcal{T}', i$} 
\State $r_1 \gets E_{R_1}.dRepID$, $r_2 \gets E_{R_2}.dRepID$
\If {$r_1 \neq r_2$}

\State $s_1 \gets \Call{Get-rep-size}{r_1, \mathcal{T}, \mathcal{T}', i}$
\State $s_2 \gets \Call{Get-rep-size}{r_2, \mathcal{T}, \mathcal{T}', i}$

\If {$(s_1 > s_2)$ \textbf{or} $(s_1 = s_2$ \textbf{and} $r_1 < r_2)$}
\State $\mathcal{T} \gets \Call{Update-rep-ET-node}{E_{R_2}, r_2, r_1, \mathcal{T}, i}$
\Else
\State $\mathcal{T} \gets \Call{Update-rep-ET-node}{E_{R_1}, r_1, r_2, \mathcal{T}, i}$
\EndIf

\EndIf
\State \Return $\mathcal{T}$
\EndProcedure \\

\Function{Get-rep-size}{$r, \mathcal{T}, \mathcal{T}', i$}
\If {\Call{HT-iskey}{$\mathcal{T}[i], r$}=TRUE}
\State $E_M \gets \Call{HT-get}{\mathcal{T}[i], r}$
\State $s \gets E_M.CG_1.size$
\Else
\State $s \gets \Call{HT-get}{\mathcal{T}'[i], r}$
\EndIf
\State \Return $s$
\EndFunction \\

\Function{Update-rep-ET-node}{$E_R, r_1, r_2, \mathcal{T}, i$}
\State $E_R.dRepID \gets r_2$
\If{\Call{HT-iskey}{$\mathcal{T}[i], r_1$} = TRUE}
\State $\mathcal{T}[i] \gets \Call{HT-delete}{\mathcal{T}[i], r_1}$
\EndIf
\State \Return $\mathcal{T}$
\EndFunction

\end{algorithmic}
\label{alg::update-rep}
\end{algorithm}
\begin{algorithm}[!tb]
\caption{Obtain the birth and death time of the nodes from $\mathcal{T}$ and save this information in $P$}
\begin{algorithmic}[1]
\Procedure{Generate-PD}{$\mathscr{M}, \mathcal{T}$}
\State $E_{M'}\gets \argmax_{(m, E_M) \in \mathcal{T}[2]}(E_M.CG_1.size)$
\State $\mathcal{T}[1] \gets (E_{M'}.CG_1.id, E_{M'})$
\For {$i =|\mathcal{T}|-1, \ldots,1$}
\State $\mathscr{M} \gets \Call{Record-death}{\mathscr{M}, \mathcal{T}, i}$
\EndFor
\State $\mathsf{M'} \gets \Call{HT-get}{\mathscr{M},E_{M'}.CG_1.id}$
\State $\mathsf{M'}.deathT \gets 1$
\State $P \gets \emptyset$
\State $\Call{Preorder}{P, \mathsf{M'}}$ 
\State \Return $\mathscr{M}, P$
\EndProcedure \\

\Function{Record-death}{$\mathscr{M}, \mathcal{T}, i$}
\State $D \gets \mathcal{T}[i+1] \backslash \mathcal{T}[i]$ 
\If {$D \neq \emptyset$}
\ForEach {$(m, E_M) \in D$}
\State $\mathsf{M} \gets \Call{HT-get}{\mathscr{M},m}$
\State $\mathsf{M}.deathT \gets i$

\ForEach {$(m', E_{M'}) \in \mathcal{T}[i]$}
\If {$i=1$ \textbf{or} $\Call{ETT-Connected}{\mathsf{M}.pCG[i-1].pET, E_{M'}}=\textrm{TRUE}$}
\State $\mathsf{M'} \gets \Call{HT-get}{\mathscr{M},m'}$
\State $\mathsf{M'}.child \gets \Call{Push-back}{\mathsf{M'}.child, \mathsf{M}}$
\Break
\EndIf

\EndFor
\EndFor 
\EndIf
\State \Return $\mathscr{M}$
\EndFunction \\

\Function{Preorder}{$P, \mathsf{M}$} 
\Comment Visit each node in the community tree through a preorder traversal starting from $\mathsf{M}$
\If {$\mathsf{M}=\textrm{NULL}$}
\State \Return
\Else
\algstore{myalg}
\end{algorithmic}
\end{algorithm}

\begin{algorithm}                     
\begin{algorithmic} [1]                   
\algrestore{myalg}
\State $P \gets P \cup \{(\mathsf{M}.deathT, \mathsf{M}.birthT)\}$
\While {$\mathsf{M}.child \neq \emptyset$}
\State $\mathsf{M'} \gets \Call{Pop-back}{\mathsf{M}.child}$
\State $\Call{Preorder}{P, \mathsf{M'}}$
\EndWhile
\EndIf
\EndFunction

\end{algorithmic}
\label{alg::generate-PD}
\end{algorithm}
In Algorithm~\ref{alg::get-neighbor-MCs}, we obtain $\mathcal{N}(M)$ for $M$ by first including the MCs from $T(u)$ for $u \in V(M)$ (lines 3-7). For adding neighboring MCs that do not start with $u \in V(M)$, it suffices to only consider the MCs present in $T(u)$ in $u \in \tilde{L}(v_j)$, $\tilde{L}(v_j) \gets L(v_j)\backslash\{v_i: v_i \in M, v_i.id < v_j.id\}$ for $j = 1, \ldots, |M|$, that contains $v_j$ (lines 8-17).

Algorithm~\ref{alg::output_mc} is straightforward. To output all the paths that contain a node $x$ in $T(v)$, we first add $x$ itself and all the nodes above $x$ to each path, and then walk down the paths below $x$ recursively in case $x$ is a shared node. 

Algorithm~\ref{alg::insert-edges} provides the implementation details of inserting tree edges. First, a CG edge $e_{\mathcal{G}}$ is initialized with endpoints and an array of pairs of null pointers. If $e_{\mathcal{G}}$ is inserted in $\mathcal{F}_i$, in addition to adding $e_{\mathcal{G}}$ to the adjacency lists of tree edges of $n_{M_1}[i]$ and $n_{M_2}[i]$, we also create two ET nodes that $e_{\mathcal{G}}.pET_1[i]$ and $e_{\mathcal{G}}.pET_2[i]$ are directed to when linking two ET trees (lines 22-24). Notice the ET nodes corresponding to the CG nodes to be connected are splayed to the roots of their respective ET trees in preparation for the linkage of two trees (lines 7-8). Meanwhile, we also need to update the representative ET node of the new tree. 

In Algorithm~\ref{alg::update-rep}, let $E_{R_1}$, $E_{R_2}$ be the respective root nodes of two trees to be connected, and $r_1$, $r_2$ be the values of the representative ID of the two trees. Since $E_{R_1}$, $E_{R_2}$ are both loop nodes, and each loop node in an ET tree is designed to carry the representative ET node's ID in the difference form of $dRepID$, we have $r_1 = E_{R_1}.dRepID$ and $r_2 = E_{R_2}.dRepID$. Due to this fact, we can directly identify the representative ET node and determine the size of the representative CG node $s_i$ for each spanning tree (lines 14-15). If $s_1 > s_2$, or $s_1 = s_2$ and $r_1 < r_2$, we add ($r_1-r_2$) to $E_{R_2}.dRepID$ before the merger to ensure that the IDs of all the loop nodes in the new ET tree rooted $E_{R_1}$ at are the same as $r_1$. We then remove the representative ET node with ID $r_2$ from $\mathcal{T}[i]$. The operation (line 9) is done the other way round in case the if condition is not true. We discuss the role of $\mathcal{T}'$ in Section~\ref{sec::update}.

In Algorithm~11, we first set the only element in $\mathcal{T}[1]$ to be the element including the oldest representative ET node in $\mathcal{T}[2]$ (lines 2-3). We then record the death time for each $\mathsf{M}$ whose corresponding element $(m, E_M)$ disappears in $\mathcal{T}[i]$, if any (lines 13-22). For each $(m', E_{M'}) \in \mathcal{T}[i]$, if $E_M$ is in the same ET tree as $E_{M'}$ (line 19), it implies the CC represented by $\mathsf{M}$ has been merged into the CC represented by $\mathsf{M'}$, i.e., $\mathsf{M}$ becomes the {\it child} of $\mathsf{M'}$ (line 21). A special case is that for the only remaining element in $\mathcal{T}[1]$, the death time of its corresponding MC, named {\it root representative MC}, is set to 1. Finally, we output the persistence diagram as an array of pairs of every representative MC's death and birth time through a preorder traversal starting from the root representative MC (lines 26-32).

\section{Community Tree Update Algorithm}
\label{sec::update}

\begin{figure}[!hb]
\centering
\includegraphics[width=\columnwidth]{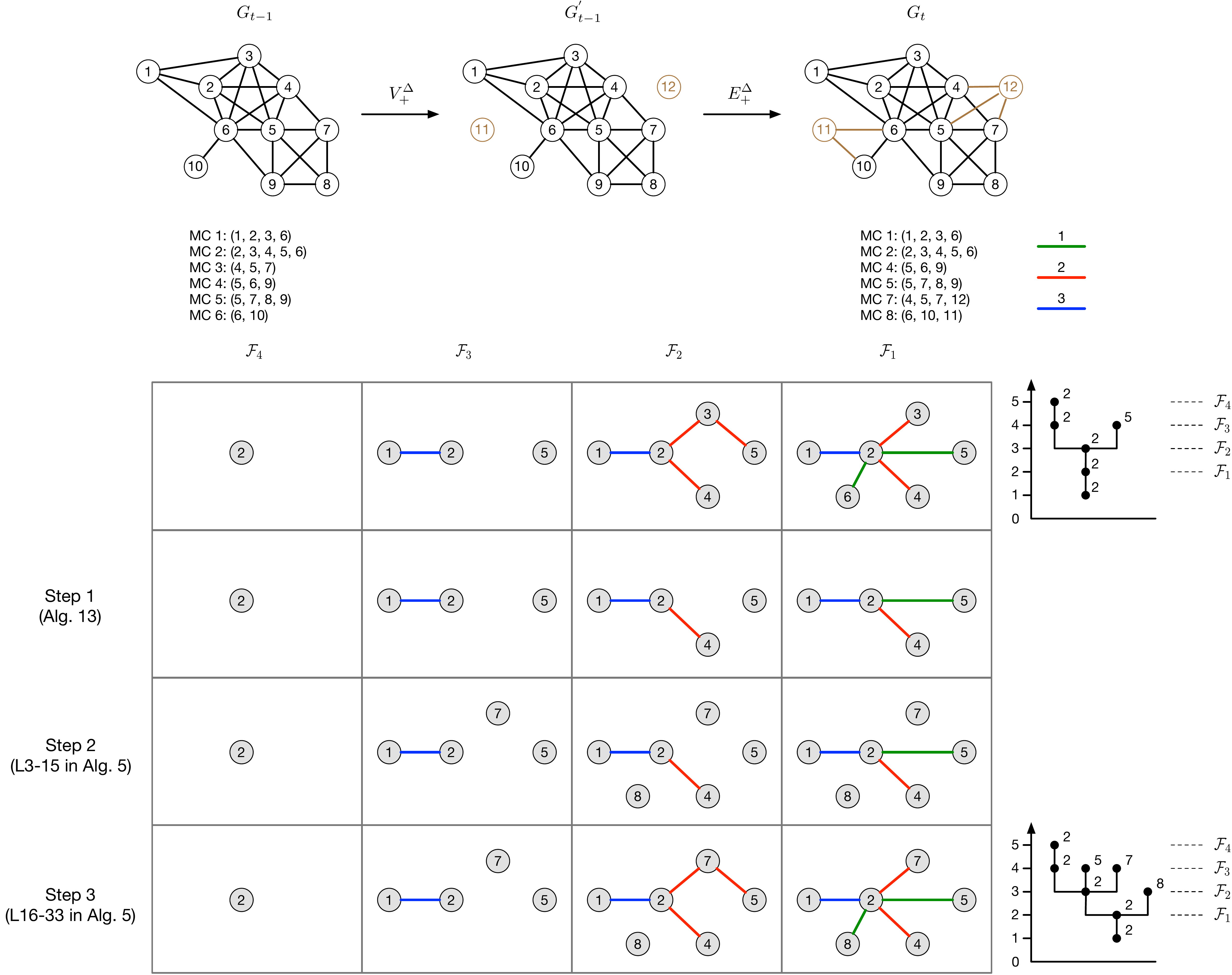}
\caption{An example of updating a community tree. $\mathcal{T}$ does not change from $G_0$ to $G'_0$ since only single vertices are added. From $G'_0$ to $G_1$, the CG nodes corresponding to MCs 3 and 6 are removed along with their incident edges before those corresponding to MCs 7 and 8 are added and connected to others. $\mathcal{T}$ is simultaneously updated due to tree edge insertions.}
\label{fig::updateCT}
\end{figure}

The variations of $G$ are simply captured through insertions of edges and vertices. As Algorithm~\ref{alg::update-CT} and Figure~\ref{fig::updateCT} demonstrate, the updates of $\mathcal{T}$ and $P$ from $G_{t-1}$ to $G_t$ are conducted through the following steps: insert each single vertex in $V^\Delta_+=V_t\backslash V_{t-1}$, where no change of $\mathcal{T}$ occurs, and then add the edges in $E^\Delta_+=E_t\backslash E_{t-1}$.
\begin{algorithm}[H]
  \caption{Update $\mathcal{T}$ after multiple insertions of vertices and edges from $G_{t-1}$ to $G_{t}$}
\begin{algorithmic}[1]
\Procedure {Update-CT}{$G_{t-1}, V^\Delta_+, E^\Delta_+, \mathscr{T}, \mathscr{M}, \mathcal{T}, m_{t-1}$}
\State $G_{t-1}^{\prime} \gets G_{t-1} + V^\Delta_+$
\For {$i = 1, \ldots, |\mathcal{T}|$}
\ForEach{$(m, E_M) \in \mathcal{T}[i]$} 
\State $\mathcal{T}'[i] \gets
\Call{HT-add}{T'[i], (m, E_M.CG_1.size)} $
\EndFor
\EndFor
\State $\mathcal{M}^{new} \gets \Call{List-new-MCs}{G_{t-1}^{\prime},E^\Delta_+}$ \Comment Algorithm 3 in~\cite{das2016incremental}
\State $\omega \gets \max_{M \in \mathcal{M}^{new}}|M|$ 
\State $\mathcal{M}^{del} \gets \Call{List-subsumed-MCs}{E^\Delta_+, \mathscr{T}, \mathcal{M}^{new}}$ 
\Comment Algorithm 4 in~\cite{das2016incremental}
\State $\mathscr{T}, \mathscr{M}, \mathcal{T} \gets \Call{Remove-MCs}{\mathcal{M}^{del},  \mathscr{T}, \mathscr{M}, \mathcal{T}}$ \Comment Algorithm~\ref{alg::remove-MCs}
\If {$\omega >|\mathcal{T}|$} \Comment Expand $\mathcal{T}$ with empty elements at the end to reach a size of $\omega$
\State $\mathcal{T} \gets \Call{Resize}{\mathcal{T}, \omega}$ 
\EndIf
\State Initialize an empty array of size $|\mathcal{T}|$ as $\mathcal{T}'$
\State $\mathscr{T}, \mathscr{M}, \mathcal{T}, m_t \gets \Call{Add-MCs}{G_t^{\prime}, \mathcal{M}^{new}, \mathscr{T}, \mathscr{M}, \mathcal{T}, \mathcal{T}', m_{t-1}}$
\State $\mathscr{M}, P \gets \Call{Generate-PD}{\mathscr{M}, \mathcal{T}}$

\State \Return $\mathscr{T}, \mathscr{M}, \mathcal{T}, P, m_{t}$
\EndProcedure
\end{algorithmic}
\label{alg::update-CT}
\end{algorithm}

\begin{algorithm}[!hb]
\caption{Update CGs after removing all the MCs in $\mathcal{M}^{del}$}
\begin{algorithmic}[1]
\Procedure{Remove-MCs}{$\mathcal{M}^{del}, \mathscr{T}, \mathscr{M}, \mathcal{T}$}
\ForEach {$M \in \mathcal{M}^{del}$}
\State $v^* \gets \argmin_{v \in M} v.id $
\State $T(v^*) \gets \Call{HT-get}{\mathscr{T}, v^*.id}$

\State $m \gets \Call{Trie-get-id}{T(v^*), M}$
\State $\mathsf{M} \gets \Call{HT-get}{\mathscr{M}, m}$
\State $\Call{Delete-CG-edges}{\mathsf{M}}$ \Comment Algorithm~\ref{alg::delete-CG-edges}
\For {$i =|M|,\ldots, 2$}
\If {\Call{HT-iskey}{$\mathcal{T}[i], m$} = TRUE}
\State $\mathcal{T}[i] \gets  \Call{HT-delete}{\mathcal{T}[i], m}$
\EndIf
\EndFor

\State $\mathscr{M} \gets \Call{HT-delete}{\mathscr{M}, m}$ 
\State $\Call{Delete}{\mathsf{M}}$ \Comment Delete $\mathsf{M}$ as an object and the corresponding CG, ET nodes at each order
\State $T(v^*) \gets \Call{Trie-remove}{T(v^*), M}$
\EndFor
\State \Return $\mathscr{T}, \mathscr{M}, \mathcal{T}$
\EndProcedure 
\end{algorithmic}
\label{alg::remove-MCs}
\end{algorithm}

\begin{algorithm}[!t]
\caption{Delete CG edges incident to the CG node corresponding to $\mathsf{M}$ at each order}
\begin{algorithmic}[1]
\Procedure{Delete-CG-edges}{$\mathsf{M}$}
\For {$i = \mathsf{M}.birthT-1, \ldots, 1$}
\While {$n_M[i].adjTrEdges \neq \emptyset$}
\State $e_{\mathcal{G}} \gets \Call{Pop-back}{n_M[i].adjTrEdges}$
\State $n_{M_1} \gets e_{\mathcal{G}}.CG_1$, $n_{M_2} \gets e_{\mathcal{G}}.CG_2$
\State $\Call{Delete-tree-edge}{n_{M_1}, n_{M_2}, e_{\mathcal{G}}, i}$
\If{$i =1$}
\State \Call{Delete}{$e_{\mathcal{G}}$}
\EndIf
\EndWhile
\EndFor
\EndProcedure \\

\Function{Delete-tree-edge}{$n_{M_1}, n_{M_2}, e_{\mathcal{G}}, i$}
\State $n_{M_1}[i].adjTrEdges \gets \Call{Remove-from-list}{n_{M_1}[i].adjTrEdges, n_{M_1}[1].id, n_{M_2}[1].id}$
\State $n_{M_2}[i].adjTrEdges \gets \Call{Remove-from-list}{n_{M_2}[i].adjTrEdges, n_{M_1}[1].id, n_{M_2}[1].id}$
\State $\Call{ETT-cut}{e_{\mathcal{G}}.pET_1[i], e_{\mathcal{G}}.pET_2[i]}$
\EndFunction \\

\Function{Remove-from-list}{$L, m_1, m_2$}
\For {$i = 1, \ldots, |L|$}
\State $n_{M_1} \gets L[i].CG_1$, $n_{M_2} \gets L[i].CG_2$
\If {$n_{M_1}[1].id = m_1$ \textbf{and} $n_{M_2}[1].id = m_2$}
\State $L \gets L\backslash L[i]$
\State \Return $L$
\EndIf
\EndFor 
\EndFunction 

\end{algorithmic}
\label{alg::delete-CG-edges}
\end{algorithm}
Algorithm~\ref{alg::update-CT} then  enumerates the new MCs $\mathcal{M}^{new}$ and the subsumed MCs\footnote{We determine if a clique is a subsumed MC through the function \Call{Trie-get-id}{}. If it returns -1, then the clique is not a subsumed MC; otherwise, it is.} $\mathcal{M}^{del}$ due to the insertion of edges in $E^\Delta_+$. The CG nodes corresponding to each $M \in \mathcal{M}^{del}$ are then removed from $\mathcal{F}_{|M|-1}, \ldots, \mathcal{F}_1$, as given in Algorithm~\ref{alg::remove-MCs}, before those corresponding to the new MCs are inserted in the CGs. The removal of a CG node implies the deletion of all the incident tree edges (Algorithm~\ref{alg::delete-CG-edges}). 
As the reverse operation of \Call{Add-tree-edge}{}, \Call{Delete-tree-edge}{} removes the tree edge $e_{\mathcal{G}}$ from the tree adjacency list of the corresponding CG nodes, and splits the ET tree into two by cutting out two arcs, $e_{\mathcal{G}}.pET_1[i]$ and $e_{\mathcal{G}}.pET_2[i]$, in the tours. After all the tree edges incident to $\mathsf{M}.pCG$ are discarded, we additionally remove $E_M$ from $T[i]$ if $M$ is a representative MC, erase $\mathsf{M}$ from $\mathscr{M}$ and delete it, along with the corresponding isolated CG and ET nodes (lines 8-12 in Algorithm~\ref{alg::remove-MCs}). 

Notice that we do not need to find the representative ET node for the separate tree after the split in Algorithm~\ref{alg::delete-CG-edges}. For each removed $M \in \mathcal{M}^{del}$, it is subsumed by a $M \in \mathcal{M}^{new}$ that will reconnect with the neighbors of the removed MC and be the candidate of the representative MC of the new CC after the reconnection. Thus, we only need to update the representative MC during \Call{Add-MCs}{}. Back to Algorithm~\ref{alg::update-rep}, the case when $r_1 = r_2$ implies that both ET trees are cut from the same tree and still retain the representative ID from the original tree. Thus, no update needs to be performed before the reconnection. We create $\mathcal{T}'$ to store the sizes of the representative MCs at each order in $G_{t-1}$ (lines 3-5 in Algorithm~\ref{alg::update-CT}). 
When $r_1 \neq r_2$, if the subsumed MC was the representative MC and removed from $\mathcal{T}[i]$ beforehand, we obtain the size of the representative MC from $\mathcal{T}'$ instead. 

\end{document}